\documentclass[a4paper,12pt]{article}
\usepackage[utf8]{inputenc}
\usepackage{graphicx,float,rotating,hyperref,slashed,verbatim,amsmath,xcolor,amssymb,amsfonts,expdlist,colortbl,youngtab}
\usepackage{mathrsfs} 
\usepackage{amsfonts}
\usepackage{halloweenmath}
\usepackage{physics}
\usepackage{amsmath}
\usepackage{physics}
\usepackage{placeins}
\usepackage{mathtools}
\usepackage{hhline}
\usepackage{comment}
\usepackage{colortbl}
\usepackage[style=numeric-comp,sorting=none,maxbibnames=9]{biblatex}
\usepackage{multicol}
\usepackage{pdflscape}
\usepackage{multirow}
\usepackage{booktabs}    
\usepackage{calc}        
\usepackage{tikz}
\usepackage[compat=1.1.0]{tikz-feynman}
\usepackage{subcaption}
\usepackage{fancyhdr}
\usepackage{cleveref}
  \bibliography{bibliography}
\hypersetup{colorlinks,bookmarksopen,bookmarksnumbered,
	linkcolor=blus,pdfstartview=FitH,urlcolor=rossos,citecolor=verde}

\usepackage{tabularx}

\definecolor{gold}{RGB}{187,161,79}
\definecolor{silver}{RGB}{192,192,192}
\definecolor{darkgreen}{RGB}{0, 130, 0}

\newcommand{\sfootnote}[1]{}
\definecolor{bluc}{cmyk}{1,1,0,0.1}
\definecolor{rossoCP3}{cmyk}{0,.88,.77,.40}
\definecolor{rosso}{cmyk}{0,1,1,0.4}
\definecolor{giallo}{cmyk}{0,.33,1,0}
\definecolor{rossos}{cmyk}{0,1,1,0.55}
\definecolor{rossoc}{cmyk}{0,1,1,0.2}
\definecolor{verdes}{cmyk}{0.92,0,0.59,0.4}

\hypersetup{colorlinks, bookmarksopen, bookmarksnumbered,
	citecolor=verdes, linkcolor=bluc, pdfstartview=FitH, urlcolor=rossos}

\newcommand{\mio}[1]{}

\definecolor{Gray}{gray}{0.95}

\usepackage{multicol}
\usepackage{color}
\definecolor{rosso}{cmyk}{0,1,1,0.4}
\definecolor{rossos}{cmyk}{0,1,1,0.55}
\definecolor{rossoc}{cmyk}{0,1,1,0.2}
\definecolor{blu}{cmyk}{1,1,0,0.3}
\definecolor{blus}{cmyk}{1,1,0,0.6}
\definecolor{bluc}{cmyk}{1,1,0,0.1}
\definecolor{verde}{cmyk}{0.92,0,0.59,0.25}
\definecolor{verdec}{cmyk}{0.92,0,0.59,0.15}
\definecolor{verdes}{cmyk}{0.92,0,0.59,0.4}

\renewcommand\&{&}

\newcommand{\GeV}{\,{\rm GeV}}

\def\circa#1{\,\raise.3ex\hbox{$#1$\kern-.75em\lower1ex\hbox{$\sim$}}\,}

\newcommand{\beq}{\begin{equation}}
	\newcommand{\eeq}{\end{equation}}

\newcommand{\bea}{\begin{eqnarray}}
	\newcommand{\eea}{\end{eqnarray}}
\newcommand{\be}{\begin{equation}}
	\newcommand{\ee}{\end{equation}}
\font\tenrsfs=rsfs10 at 12pt
\font\sevenrsfs=rsfs7 at 10 pt
\font\fiversfs=rsfs5
\newfam\rsfsfam
\textfont\rsfsfam=\tenrsfs
\scriptfont\rsfsfam=\sevenrsfs
\scriptscriptfont\rsfsfam=\fiversfs

\makeatletter

\def\hhref#1{\href{http://arxiv.org/abs/#1}{arXiv:#1}} 

\newcommand{\doi}[1]{\href{http://dx.doi.org/#1}{[doi]}}

\def\hhref#1{\href{http://arxiv.org/abs/#1}{arXiv:#1}}

\def\art{\@ifnextchar[{\eart}{\oart}}
\def\eart[#1]#2#3#4#5#6{{\rm #2}, {\em #3 \bf #4} {\rm (#6) #5} ({\em #1})}

\def\article{\@ifnextchar[{\earticle}{\oarticle}}
\def\oarticle#1#2#3#4#5#6{{\rm #1}, {\em ``#6''}, {\rm #2 #3 (#5) #4}}
\def\earticle[#1]#2#3#4#5#6#7{{\rm #2}, {\em ``#7''}, {\rm #3 #4 (#6) #5}  [\hhref{#1}]}
\def\hepart[#1]#2{{\rm #2, \em#1}}
\def\heparticle[#1]#2#3{#2, {\em ``#3''} [\hhref{#1}]}

\newcounter{alphaequation}[equation]
\def\thealphaequation{\theequation\hbox to
	0.6em{\hfil\alph{alphaequation}\hfil}}
\def\eqnsystem#1{
	\def\@eqnnum{{\rm (\thealphaequation)}}
	\def\@@eqncr{\let\@tempa\relax \ifcase\@eqcnt \def\@tempa{& & &} \or
		\def\@tempa{& &}\or \def\@tempa{&}\fi\@tempa
		\if@eqnsw\@eqnnum\refstepcounter{alphaequation}\fi
		\global\@eqnswtrue\global\@eqcnt=0\cr}
	\refstepcounter{equation} \let\@currentlabel\theequation \def\@tempb{#1}
	\ifx\@tempb\empty\else\label{#1}\fi
	\refstepcounter{alphaequation}
	\let\@currentlabel\thealphaequation
	\global\@eqnswtrue\global\@eqcnt=0 \tabskip\@centering\let\\=\@eqncr
	$$\halign to \displaywidth\bgroup \@eqnsel\hskip\@centering
	$\displaystyle\tabskip\z@{##}$&\global\@eqcnt\@ne
	\hskip2\arraycolsep\hfil${##}$\hfil& \global\@eqcnt\tw@\hskip2\arraycolsep
	$\displaystyle\tabskip\z@{##}$\hfil
	\tabskip\@centering&\llap{##}\tabskip\z@\cr}
\def\endeqnsystem{\@@eqncr\egroup$$\global\@ignoretrue} \makeatother

\newcommand{\SU}{\,{\rm SU}}

\definecolor{fiorentina}{rgb}{.5,0,.5}

\usepackage{tikz-feynman}

\let\oldheadrule\headrule
\renewcommand{\headrule}{\color{white}\oldheadrule}
\begin{document}

	\begin{center}
		\boldmath
		
		{\textbf{\LARGE\color{rossoCP3}
		 New Physics Pathways from $B$ Processes
			}}
		\unboldmath
		
		\bigskip\bigskip

	 	{ Alessandra D'Alise$^1$, Giuseppe~Fabiano$^1$, Domenico~Frattulillo$^1$,  Davide~Iacobacci$^{1}$,   Francesco~Sannino$^{1,2,3}$, Pietro~Santorelli$^1$, Natascia~Vignaroli$^{1,4}$\\[5mm]}

\small{$^1$ Dipartimento di Fisica ``E. Pancini", Università di Napoli Federico II - INFN sezione di Napoli, Complesso Universitario di Monte S. Angelo Edificio 6, via Cintia, 80126 Napoli, Italy}\\
\small{$^2$ Scuola Superiore Meridionale, Largo S. Marcellino, 10, 80138 Napoli NA, Italy}\\
\small{$^3$  Quantum  Theory Center ($\hbar$QTC), Danish-IAS, IMADA, Southern Denmark Univ., Campusvej 55, 5230 Odense M, Denmark}\\
\small{$^4$  Dipartimento di Matematica e Fisica ``E. De Giorgi", Universit\`a del Salento and INFN Lecce, \\ via per Arnesano 73100 Lecce, Italy}\\

		\bigskip\bigskip
		
		\thispagestyle{empty}\large
		{\bf\color{blus} Abstract}
		\begin{quote} 
		\normalsize

We re-consider recent measures of $R_{K}$ and  $R_{K^*}$, now compatible with the Standard Model expectations,  as well as the results for the process $\text{BR}(B_s \rightarrow \mu^+ \mu^-)$ alongside earlier determinations of $R_{D^{(\ast)}}$ and $\text{BR}(B_c \rightarrow \tau \nu)$.  We  provide analytic constraints on the associated Wilson coefficients in both the $b \to s$ and the $b \to c$ sectors. These allow us to estimate the scale of potential New Physics for generic extensions of the Standard Model. We then use the results to constrain the leptoquark landscape. 
 
		\end{quote}
		\thispagestyle{empty}
	\end{center}
	
	\setcounter{page}{1}
	\setcounter{footnote}{0}
	
	
\thispagestyle{fancy}

	\newpage
	\tableofcontents

	\newpage
 
	\section{Introduction}

Experimental tests at colliders of the  Standard Model have crowned it as the  golden standard of our understanding of fundamental interactions. However, the model is far from being satisfactory. From a theoretical standpoint, there is no  underlying explanation for the gauge structure nor why we observe three matter generations, the Higgs sector is unnatural and the theory, in absence of gravity, develops an UV cutoff (Landau Pole)  and there is  yet no consensus on a consistent theory of quantum gravity and particle physics. On the experimental front, the Standard Model fails to explain the matter-anti matter asymmetry, it lacks a candidate for dark matter and does not address the origin of neutrino masses. It is therefore fair to say that we have so far established an excellent "Effective Model" of fundamental interactions and that sometime (hopefully) soon even collider experiments will provide clear indications on which extension should be considered as the new Standard Model. 

Of course, any extension of the Standard Model will have to reproduce its successes at lower energies, and one can use this to narrow the pathways for new physics models. In our recent review \cite{DAlise:2022ypp} we laid the foundations for a general investigation of models of new physics stemming from experimental test of the Standard Model, from B physics to lepton $g-2$.   

Given the  experimental updates such as the ones about  lepton-flavor universality \cite{LHCb:2022qnv,LHCb:2022zom}  we re-analyze their impact on possible deviations from the Standard Model     while also  taking into account the anomalies in the $R_{D^{\ast}}$ measurements \cite{HeavyFlavorAveragingGroup:2022wzx} that we had omitted in \cite{DAlise:2022ypp}. Additionally we consider also the still significant deviations from the SM predictions in rare $B$ meson decays~\cite{LHCb:2014cxe,BELLE:2019xld,LHCb:2017avl,LHCb:2021trn,LHCb:2021awg,LHCb:2021zwz,LHCb:2020gog,Belle:2019oag,ATLAS:2018cur,CMS:2019qnb,LHCb:2016ykl,BaBar:2013qry,Belle:2005fli,CMS:2017ivg,ATLAS:2017dlm}. 
 In particular, we will consider the leptoquark landscape which has been often used in the literature to discuss new physics in the flavour sector either as elementary extensions of the Standard Model or as effective descriptions of some underlying composite dynamics. As we shall see, our estimates can be used to guide searches of new physics at present and future colliders.

The work is organized as follows. In Section~\ref{bsFConstrains} and \ref{btoc} we re-analyse the experimental results and their impact on the associated theoretical effective description in terms of Wilson coefficients for $b\rightarrow c$ and $b\rightarrow s$ observables. An analytic study is performed for those observables that are less affected by hadronic physics contamination which are $R_{K^{(\ast)}}$,  $R_{D^{(\ast)}}$, $\text{BR}(B_s \to \mu^+ \mu^-)$ and $\text{BR}(B_c \to \tau \nu)$. We further compare the analytic results with numerical fits using FLAVIO \cite{Straub:2018kue} consolidating our results. Having gained information on the maximum deviation allowed by the Wilson coefficients we link them to time-honoured underlying models in Section~\ref{sec:th} such as the $Z'$ and different type of scalar and vector leptoquarks (LQ). Assuming our focus on  models of new physics (NP) in which the new couplings are not too finely tuned, we show that to simultaneously accommodate Standard Model deviations in $R_{D^{(\ast)}}$, while respecting the constraints coming from $R_{K^{(\ast)}}$, one can employ the weak singlet scalar leptoquark $S_1$, with a mass in the TeV range.\\
Additionally, our analysis suggests that alternative leptoquark models require large fine-tuned couplings to effectively account for both $b \to s$ and $b \to c$ data. Subsequently, we furnish bounds on the possible NP scales.

	\section{\texorpdfstring{$b \to s$} ~ Flavour Constraints}
\label{bsFConstrains}

To explore NP emerging around the electroweak energy scale affecting the bottom to strange transitions involving leptons one is led to introduce the following class of effective operators	\beq {\cal O}_{b_X \ell_Y} = (\bar s \gamma_\mu P_X b)(\bar \ell \gamma_\mu P_Y \ell) \ ,\eeq
 which can be written as $\SU(2)_L$-invariant operators. A more general discussion can be found in ~\cite{Aebischer:2017gaw,Aebischer:2018bkb,Alonso:2014csa}.
 These operators are  incorporated  in the following effective Hamiltonian 
	\beq \mathscr{H}_{\rm eff}
	=  - V_{tb} V_{ts}^*  \frac{\alpha_{\rm em}}{4\pi v^2}
	\sum_{\ell, X, Y} C_{b_X \ell_Y}  {\cal O}_{b_X \ell_Y}\, + \mathrm{h.c.} \, , \eeq
	where the sum runs over leptons $\ell =   \{e,\mu , \tau\}$ and over their chiralities $X,Y = \{L,R\}$. Additionally, it is convenient to define dimensionless coefficients $C_I$ related to the dimensionful $c_I$ coefficients appearing in the equivalent Lagrangian formulation 
	\begin{equation}
	\mathscr{L}_{\rm eff}=
	\sum_{\ell, X, Y} c_{b_X \ell_Y}  {\cal O}_{b_X \ell_Y} \ , \; \text{with} \quad c_I = V_{tb} V_{ts}^*  \frac{\alpha_{\rm em}}{4\pi v^2} C_I  \ , \eeq

	where $V_{ts}= - 0.0412 \pm 0.0006$  and 
	$v= (2\sqrt{2}G_{\rm F})^{-1/2} = 174\, \GeV$ is the Higgs vacuum expectation value and $G_{\rm F}$ the Fermi constant.

\subsection{Applications to \texorpdfstring{$R_{K^{(*)}}$}{RK(*)} and \texorpdfstring{$\text{BR}(B_s \to \mu^+ \mu^-)$}{BR(Bs to mu+ mu-)}}

\label{sec:bsth_analysis}
As main applications of the formalism above we start by considering the well known   $R_K$ and $R_{K^*}$ ratios \cite{BELLE:2019xld,LHCb:2021trn,LHCb:2017avl,Belle:2019oag} 
\beq \label{eq:RKRK*}
		R_K =\frac{{\rm BR} \left ( B^+ \to K^+ \mu ^+ \mu ^-\right )}{{\rm BR} \left ( B^+ \to K^+ e^+ e^-\right )}, \quad 	R_{K^*} =\frac{{\rm BR} \left ( B \to K^* \mu ^+ \mu ^-\right )}{{\rm BR} \left ( B \to K^* e^+ e^-\right )} . 
		\eeq
These quantities are known to be excellent  tests of lepton flavour universality since they are constructed to reduce QCD-related uncertainties \cite{Hiller:2003js}. Therefore,  together with   the   $B_s \to \mu^+ \mu^-$ branching ratio  \cite{LHCb:2021awg,ATLAS:2018cur,CMS:2019qnb} are hadronic insensitive \cite{DAmico:2017mtc} quantities.  Here,  `hadronic insensitive' (HI) \cite{DAmico:2017mtc}  refers to the fact that the related observables have at most few percent QCD-induced theoretical errors.  
 
The latest experimental results for $R_{K^{(*)}}$  read \cite{LHCb:2022qnv,LHCb:2022zom}: 
\begin{align}
&R_{K}= 0.994^{ +0.090}_{-0.082}({\rm{stat}})^{ +0.029}_{-0.027}({\rm{syst}}) {\qq{with}} q^2 \in [0.1,1.1] \, \rm{GeV^2} \ ,  \\
&R_{K^*}= 0.927^{ +0.093}_{-0.087}({\rm{stat}})^{ +0.036}_{-0.035}({\rm{syst}}) {\qq{with}} q^2 \in [0.1,1.1] \, \rm{GeV^2} \ , \\
&R_{K}= 0.949^{ +0.042}_{-0.041}({\rm{stat}})^{ +0.022}_{-0.022}({\rm{syst}}) {\qq{with}} q^2 \in [1.1,6] \, \rm{GeV^2} \ , \\
&R_{K^*}= 1.027^{ +0.072}_{-0.068}({\rm{stat}})^{ +0.027}_{-0.026}({\rm{syst}}) {\qq{with}} q^2 \in [1.1,6] \, \rm{GeV^2} \ .
\label{expRKRKs}
\end{align}
At high $q^2$ one can neglect the lepton masses and the formulae for $R_{K}$ and $R_{K^*}$ simplify to\footnote{
It is worth to mention that the analytic formulas given in Eqs. (\ref{eq:RKTheory}, and  \ref{eq:RKsTheory}) provide a less precise approximation for the lower bin $[0.1,1.1]$. Nevertheless, it is important to observe that measurements of the lower $q^2$ interval are associated with a higher level of uncertainty compared to the higher bin.}
\begin{equation}  
 \label{eq:RKTheory}
	R_K 
	= \frac{|C_{b_{L+R} \mu_{L-R}}|^2  + |C_{b_{L+R} \mu_{L+R}}|^2}{|C_{b_{L+R} e_{L-R}}|^2  + |C_{b_{L+R} e_{L+R}}|^2} \ ,
\end{equation}
 \begin{equation}
\label{eq:RKsTheory}
R_{K^*}
	= \frac{(1-p) (|C_{b_{L+R} \mu_{L-R}}|^2  + |C_{b_{L+R} \mu_{L+R}}|^2 ) + p \left( |C_{b_{L-R} \mu_{L-R}}|^2  + |C_{b_{L-R} \mu_{L+R}}|^2 \right) }
	{(1-p)( |C_{b_{L+R} e_{L-R}}|^2  + |C_{b_{L+R} e_{L+R}}|^2 ) + p \left( |C_{b_{L-R}e_{L-R}}|^2  + |C_{b_{L-R}e_{L+R}}|^2 \right)} \ ,
\end{equation}
	where $p\approx 0.86$ is the ``polarisation fraction"~\cite{Bobeth:2008ij,Hambrock:2013zya,Hiller:2014ula}, that is defined as
	\begin{equation}
		p =\frac{g_0 +g_{\parallel}}{g_0 +g_{\parallel}+g_{\perp}} \, .
\end{equation}
The $g_i$ are the contributions to the decay rate (integrated over the intermediate bin) of the different helicities of the $K^*$. The index $i$ distinguishes the various helicities: longitudinal ($i=0$), parallel ($i=\parallel$) and perpendicular ($i=\perp$).\\
We also use the notation \cite{DAmico:2017mtc,DAlise:2022ypp}:
\begin{equation}
	C_{b_{L\pm R} \ell_{Y}} \equiv  C_{b_L \ell_Y} \pm C_{b_R\ell_Y}, \quad
	C_{b_{L+R} \ell_{L\pm R}} \equiv C_{b_L \ell_L}+ C_{b_R \ell_L} \pm C_{b_L \ell_R} \pm C_{b_R \ell_R} \ .
\end{equation}

The coefficients $C_{b_X\ell_Y}$ are linked to $C_9$ and $C_{10}$ via
	\begin{equation}
	 2C_9 = {C_{b_L \ell_{L+R}}}, \quad  2C_{10}=-{C_{b_L \ell_{L-R}}}, \quad 2C'_9 = {C_{b_R \ell_{L+R}}}, \quad  2C'_{10}=-{C_{b_R \ell_{L-R}}} \ .
	 \end{equation}
We split the coefficients into the sum of the Standard Model contribution and the beyond Standard Model one via
\begin{equation}
C_{b_X\ell_Y}=C_{b_X\ell_Y}^{SM}+C_{b_X\ell_Y}^{BSM} \ .
\end{equation} 
We use the Standard Model values given in \cite{DAmico:2017mtc,DAlise:2022ypp} to numerically specify expressions \eqref{eq:RKTheory}-\eqref{eq:RKsTheory}: $C^{SM}_{b_L \ell_L} = 8.64$ and $C^{ SM}_{b_L \ell_R} = -0.18$, while $C^{ SM}_{b_R \ell_X} = 0$ for $X=L,R$ . \\
 We assume that possible NP corrections occur in the muonic Wilson coefficients (i.e. $C^{BSM}_{b_Xe_Y}=0$), implying Lepton Flavor Universality Violation (LFUV). Of course, they could appear in other sectors as well but experimentally muons are easier to monitor and therefore we focus on this possibility.  Figure~\ref{fig:singlecoeff} is the update of Figure 3  in \cite{DAlise:2022ypp,DAmico:2017mtc} and describes the effects on $R_{K^*}$ and $R_K$ obtained switching on one NP Wilson coefficient at a time. 
 The recent experimental measurements are exemplified at one and two sigma by the green $q^2$ bin $[1.1,6] \,\rm{GeV^2}$ and dark-gray $q^2$ bin $[0.1,1.1] \,\rm{GeV^2}$ crosses. 
\begin{figure}[h!]
    \centering
\includegraphics[scale=0.5]{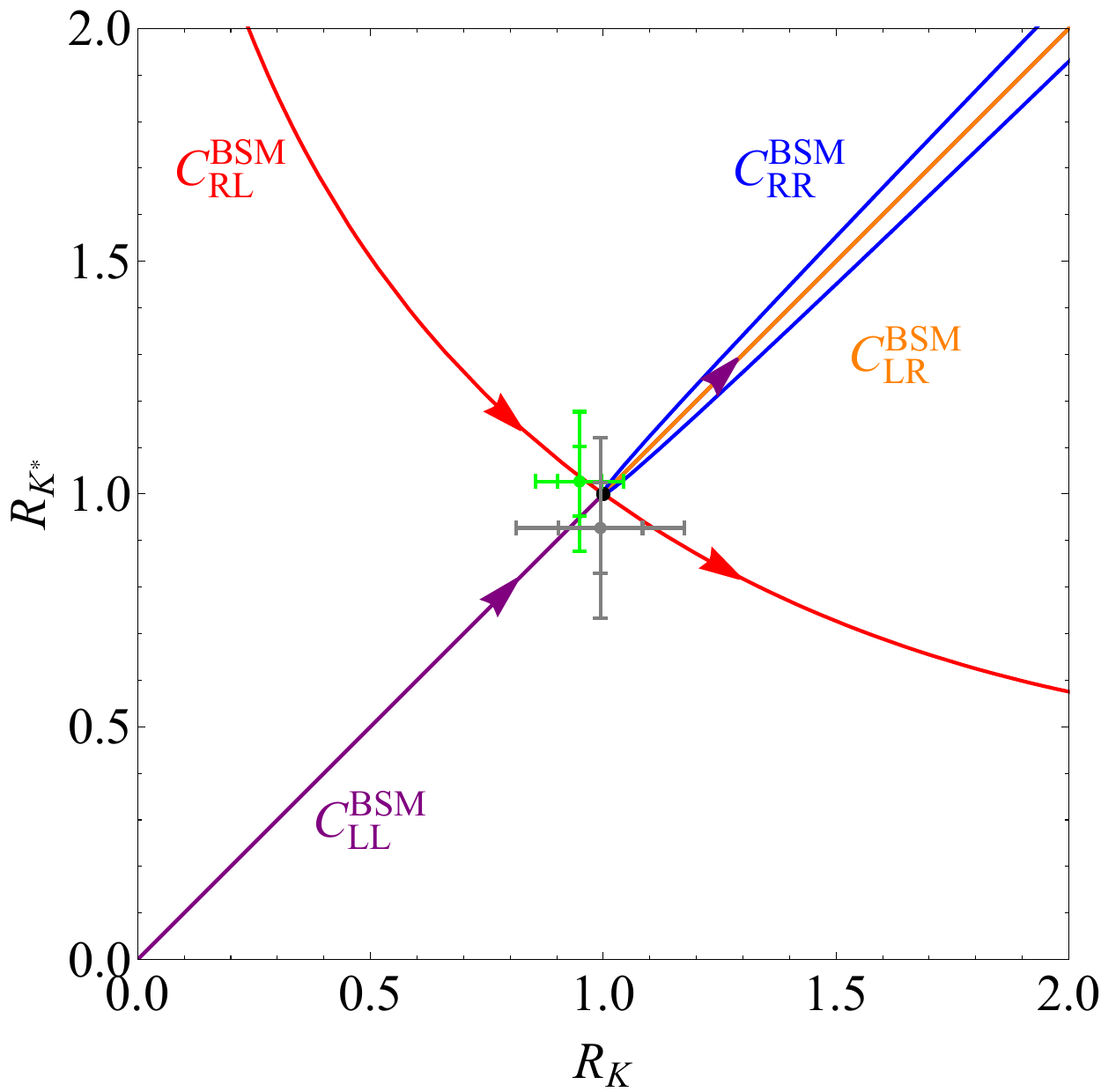}
\caption{Quantitative behaviour of $R_{K}$ vs $R_{K^*}$ as in \eqref{eq:RKTheory}-\eqref{eq:RKsTheory}, normalised by their SM value, obtained by switching on one NP coefficient at a time. The experimental values for the $[1.1,6] \, \rm{GeV^2}$ bin (in green) and for the $[0.1,1.1] \, \rm{GeV^2}$ bin (in gray) are shown with $1\sigma$ and $2\sigma$ error bars. The experimental values are compatible with the SM in the $2$-sigma range, signaling values of the NP coefficients close to 0.}
 \label{fig:singlecoeff}
\end{figure}
 Substituting in \eqref{eq:RKTheory}-\eqref{eq:RKsTheory} the SM values for the Wilson coefficients, we obtain
\begin{equation}
    \begin{aligned}
    R_K=&1+0.231[Re(C_{b_L\mu_L}^{BSM})+Re(C_{b_R\mu_L}^{BSM})]-0.005[Re(C_{b_L\mu_R}^{BSM})+Re(C_{b_R\mu_R}^{BSM})]+\\&+
    0.013\{[Re(C_{b_L\mu_L}^{BSM})+Re(C_{b_R\mu_L}^{BSM})]^2+
       [Re(C_{b_L\mu_R}^{BSM})+Re(C_{b_R\mu_R}^{BSM})]^2\}  \ , \\
   R_{K^*}-R_K=&-0.398 Re(C_{b_R\mu_L}^{BSM})+0.008 Re(C_{b_R\mu_R}^{BSM})\\&-0.046[Re(C_{b_L\mu_L}^{BSM})Re(C_{b_R\mu_L}^{BSM}))
    +Re(C_{b_L\mu_R}^{BSM})Re(C_{b_R \mu_R}^{BSM})] \ .
    \end{aligned}
\end{equation}
Another relevant clean observable for studying constraints on the NP coefficient is the branching ration ${\rm{BR}}(B_s\rightarrow \mu^+ \mu^-)$, which in terms of NP coefficients reads
	\begin{equation}
	    \label{eq:BSmumu} \textrm{BR}(B_s \to \mu^+ \mu^-)= \textrm{BR}(B_s \to \mu^+ \mu^-) _{\rm SM}
		\bigg|\frac{C_{b_{L-R}\mu_{L-R}}}{C^{\rm SM}_{b_{L-R}\mu_{L-R}}}\bigg|^2 \ ,
  \end{equation}
with $\textrm{BR}(B_s \to \mu^+ \mu^-)_{\textrm{SM}} = (3.66\pm 0.14) \times 10^{-9}$. Following the most recent measurement by the CMS collaboration \cite{CMS:2022mgd}, the experimental value, as reported in \cite{Allanach:2022iod}, for the branching ratio $\textrm{BR}(B_s \to \mu^+ \mu^-)_{\textrm{exp}}$ is now $(3.28\pm 0.26) \times 10^{-9}$. This measurement is now even closer to the Standard Model prediction. Inserting the numerical values for the SM Wilson coefficients the above becomes 
        	\beq
        	\label{eq:SManalyticalBs}
        		\begin{split} 
     \frac{\textrm{BR}(B_s \to \mu^+ \mu^-) }{\textrm{BR}(B_s \to \mu^+ \mu^-) _{\rm SM}} 	&=1+0.227[Re(C^{BSM}_{b_L\mu_L})+Re(C^{BSM}_{b_R\mu_R})-Re(C^{BSM}_{b_L\mu_R})- Re(C^{BSM}_{b_R\mu_L})]+\\ &+0.013[Re(C^{BSM}_{b_L\mu_L})+Re(C^{BSM}_{b_R\mu_R})-Re(C^{BSM}_{b_L\mu_R})- Re(C^{BSM}_{b_R\mu_L})]^2
    	\end{split}
	\eeq

As a preliminary step, we consider the $2$-$\sigma$ interval for the experimental values presented above in order to obtain constraints on the Wilson coefficients.  As mentioned earlier the values of $R_K$ and $R_{K^*}$ in the lower $q^2$ interval are associated with a higher level of uncertainty compared to the higher bin. As a result the relevant constraint stems from the $q^2$ bin $[1.1,6] \rm{GeV^2}$, as well as from the ${\rm{BR}}(B_s\to \mu^+\mu^-)$. \\
When switching on one real Wilson coefficient at a time, our analytical analysis, at the $2$-$\sigma$ level, yields the following constraints:
\begin{align}
\label{Constraintbtos1}
    &C_{b_L \mu_L}^{\rm BSM} \in [-0.53,0.19] \ , \\
    \label{Constraintbtos2}
    &C_{b_L \mu_R}^{\rm BSM} \in [-0.46,1.51] \ , \\ 
    \label{Constraintbtos3}
    &C_{b_R \mu_L}^{\rm BSM} \in [-0.46,0.19] \ , \\
    \label{Constraintbtos4}
    &C_{b_R \mu_R}^{\rm BSM}\in [-1.51,0.46] \ .
\end{align}
These intervals provide a first idea of the acceptable ranges for the associated Wilson coefficients, taking into account uncertainties at $2$-$\sigma$ confidence level. We have also considered the imaginary parts of the Wilson coefficients, but we will concentrate on the real parts since these are preferred by the data. 

\subsection{Fit Analysis}
So far we examined the essential features of the HI observables, laying the groundwork for our expectations in the comparisons with experimental data.\\
Building upon the analysis presented in \cite{DAmico:2017mtc,DAlise:2022ypp}, this section utilises the {\sc Flavio} toolkit~\cite{Straub:2018kue} to revise the outcomes of the fitting procedure, refining the values of the Wilson NP coefficients to optimally align with the available dataset.
In the initial phase of our analysis, we concentrate exclusively on the HI set of observables. Following this, we estimate the influence of the 'Hadronic Sensitive' (HS) observables and subsequently incorporate them into a unified global fitting procedure.\\
In \cref{tab:Fit2023bs}, we provide a concise overview of the updated outcomes obtained from the fitting procedure, focusing on the scenario where individual Wilson coefficients are turned on one at a time. To ensure a comprehensive update of the findings presented in \cite{DAlise:2022ypp}, we extend our analysis to include situations where NP effects are assumed in the electron sector only. \\

The obtained results align with the analytic investigation presented in \cref{sec:bsth_analysis}. The overall deviation from the SM is significantly mitigated with respect to previous results \cite{DAmico:2017mtc,Ghosh:2014awa,Altmannshofer:2014rta,Descotes-Genon:2015uva,Altmannshofer:2013foa,Hurth:2013ssa,Hiller:2014yaa,Alonso:2014csa,Hurth:2014vma,Ciuchini:2017mik,Capdevila:2017bsm,Altmannshofer:2017fio,Alguero:2019ptt,Aebischer:2019mlg,Ciuchini:2020gvn,DAlise:2022ypp}. This is to be expected because  the recent experimental measurements align well with the SM predictions.  In fact, the muon Wilson coefficients are now in agreement with the SM within a $1.2 \sigma$ confidence level. As for the muon case, the electron sector fit showes compatibility with the SM with a significance of $0.7 \sigma$. For completeness, we show in \cref{tab:Fit2023bsaxial} the results of $1$-parameter fits in the vector-axial basis.\\
In summary, when focusing the analysis on the specific subset of HI observables $R_K$, $R_{K^*}$ and ${\rm BR}(B_s \to \mu^+\mu^-)$, the overall results indicate a compatibility with the Standard Model within nearly a $1 \sigma$ level.\\
Nevertheless, a significant deviation in the muon sector persists, originating from the HS observables.  In the electron sector, however, one observes  a reduced deviation from the SM  which now hovers around $1\sigma$. This is to be expected since the majority of HS observables deviating from the SM involve muons. \\ 
We finally combine HS and HI observables in a global fit. When considering the single coefficient turned on, in the muon sector, we  conclude that the results favor a deviation in the SM in $C^{\rm BSM}_{b_L\mu_L}$ as well as $C^{\rm BSM}_{b_L\mu_R}$, with a larger significance for the left-handed muon coefficient than the right-handed one. Meanwhile, we observe that $C^{\rm BSM}_{b_R\mu_L}$ and $C^{\rm BSM}_{b_R\mu_R}$ remain compatible with the SM value, as do all the electronic coefficients.

	\begin{table}[pt]
		\begin{center}
			\begin{tabular}{crrrrrrrrrc}
                \toprule
				& \multicolumn{9}{c}{\textbf{New physics in the muon sector (Chiral basis)}} &    \\
				\cmidrule{2-10}
				& \multicolumn{3}{c}{Best-fit} & \multicolumn{3}{c}{1-$\sigma$ range} &
				\multicolumn{3}{c}{$\sqrt{\chi^2_{\rm SM} - \chi^2_{\rm best}}$}   &\\ \cmidrule{2-10}
				& \emph{HI}  & \emph{HS} & \emph{all}  & \emph{HI}  & \emph{HS} & \emph{all} & \emph{HI} & \emph{HS} & \emph{all} &   \\ \cmidrule{2-10}
				\multirow{2}{*}{$C_{b_L \mu_L}^{\rm BSM}$} &  \multirow{2}{*}{$-0.15$} & \multirow{2}{*}{$-1.31$} &  \multirow{2}{*}{$-0.33$}  & $-0.05$ & $-1.05$  & $-0.24$ & \multirow{2}{*}{$1.1$} &  \multirow{2}{*}{$4.1$} & \multirow{2}{*}{$2.8$}  & \\
				& & & & $-0.25$ & $-1.56$ & $-0.42$ & & & &\\  \cmidrule{2-10}
				\multirow{2}{*}{$C_{b_L \mu_R}^{\rm BSM}$} &  \multirow{2}{*}{$0.40$} & \multirow{2}{*}{$-0.66$} &  \multirow{2}{*}{$-0.25$}  & $0.64$ & $-0.47$  & $-0.10$ & \multirow{2}{*}{$1.2$} &  \multirow{2}{*}{$2.6$} & \multirow{2}{*}{$1.7$} & \\
				& & & & $0.16$ & $-0.85$ & $-0.40$ & & && \\  \cmidrule{2-10}
				\multirow{2}{*}{$C_{b_R \mu_L}^{\rm BSM}$} &  \multirow{2}{*}{$-0.05$} & \multirow{2}{*}{$0.08$} &  \multirow{2}{*}{$-0.04$}  & $0.05$ & $0.19$  & $0.04$ & \multirow{2}{*}{$0.3$} &  \multirow{2}{*}{$0.5$} & \multirow{2}{*}{$0.3$}  & \\
				& & & & $-0.15$ & $-0.03$ & $-0.12$ & && & \\  \cmidrule{2-10}
				\multirow{2}{*}{$C_{b_R \mu_R}^{\rm BSM}$} &  \multirow{2}{*}{$-0.38$} & \multirow{2}{*}{$0.30$} &  \multirow{2}{*}{$0.05$}  & $-0.13$ & $0.52$  & $0.20$ & \multirow{2}{*}{$1.1$} &  \multirow{2}{*}{$1.6$} & \multirow{2}{*}{$0.2$} &  \\
				& & & & $-0.63$ & $0.18$ & $-0.10$ & && & \\ 
				\toprule
				& \multicolumn{9}{c}{\textbf{New physics in the electron sector (Chiral basis)}} &    \\
				\cmidrule{2-10}
				& \multicolumn{3}{c}{Best-fit} & \multicolumn{3}{c}{1-$\sigma$ range} &
				\multicolumn{3}{c}{$\sqrt{\chi^2_{\rm SM} - \chi^2_{\rm best}}$}   &\\ \cmidrule{2-10}
				& \emph{HI}  & \emph{HS} & \emph{all}  & \emph{HI}  & \emph{HS} & \emph{all} & \emph{HI} & \emph{HS} & \emph{all} &   \\ \cmidrule{2-10}
				\multirow{2}{*}{$C_{b_L e_L}^{\rm BSM}$} &  \multirow{2}{*}{$0.10$} & \multirow{2}{*}{$0.94$} &  \multirow{2}{*}{$0.14$}  & $0.21$ & $1.45$  & $0.25$ & \multirow{2}{*}{$0.7$} &  \multirow{2}{*}{$1.2$} & \multirow{2}{*}{$0.9$}  & \multirow{2}{*}{ }\\
				& & & & $-0.01$ & $0.43$ & $0.03$ & & & &\\  \cmidrule{2-10}
				\multirow{2}{*}{$C_{b_L e_R}^{\rm BSM}$} &  \multirow{2}{*}{$-0.17$} & \multirow{2}{*}{$-2.71$} &  \multirow{2}{*}{$-0.70$}  & $1.03$ & $-1.03$  & $-0.11$ & \multirow{2}{*}{$0.1$} &  \multirow{2}{*}{$1.3$} & \multirow{2}{*}{$0.6$} & \multirow{2}{*}{ }\\
				& & & & $-1.37$ & $-1.73$ & $-1.29$ & & && \\  \cmidrule{2-10}
				\multirow{2}{*}{$C_{b_R e_L}^{\rm BSM}$} &  \multirow{2}{*}{$0.14$} & \multirow{2}{*}{$-3.87$} &  \multirow{2}{*}{$0.15$}  & $0.25$ & $-2.86$  & $0.26$ & \multirow{2}{*}{$0.9$} &  \multirow{2}{*}{$1.4$} & \multirow{2}{*}{$1.0$}  & \\
				& & & & $0.03$ & $-4.88$ & $0.04$ & && & \\  \cmidrule{2-10}
				\multirow{2}{*}{$C_{b_R e_R}^{\rm BSM}$} &  \multirow{2}{*}{$-1.22$} & \multirow{2}{*}{$-3.94$} &  \multirow{2}{*}{$-1.43$}  & $-0.59$ & $-2.92$  & $0.08$ & \multirow{2}{*}{$0.9$} &  \multirow{2}{*}{$1.4$} & \multirow{2}{*}{$1.1$} &   \multirow{2}{*}{ }\\
				& & & & $-1.85$ & $-4.96$ & $-2.94$ & && & \\  \bottomrule
			\end{tabular}
		\end{center}
		\caption{\em Best fits turning on a single operator at a time in the chiral basis, using the `hadronic insensitive' observables `HI',  the `hadronic sensitive' observables `HS', or all the observables `all'.
		The full list of observables can be found in Appendix~\ref{appbtoc}.
			\label{tab:Fit2023bs}}
	\end{table}

\begin{table}[pt]
		\begin{center}
			\begin{tabular}{crrrrrrrrrc}
                \toprule
				& \multicolumn{9}{c}{\textbf{New physics in the muon sector (Vector Axial basis)}} &    \\
				\cmidrule{2-10}
				& \multicolumn{3}{c}{Best-fit} & \multicolumn{3}{c}{1-$\sigma$ range} &
				\multicolumn{3}{c}{$\sqrt{\chi^2_{\rm SM} - \chi^2_{\rm best}}$}   &\\ \cmidrule{2-10}
				& \emph{HI}  & \emph{HS} & \emph{all}  & \emph{HI}  & \emph{HS} & \emph{all} & \emph{HI} & \emph{HS} & \emph{all} &   \\ \cmidrule{2-10}
				\multirow{2}{*}{$C_{9,\mu}^{\rm BSM}$} &  \multirow{2}{*}{$-0.12$} & \multirow{2}{*}{$-0.94$} &  \multirow{2}{*}{$-0.49$}  & $-0.24$ & $-0.82$  & $-0.39$ & \multirow{2}{*}{$0.7$} &  \multirow{2}{*}{$4.9$} & \multirow{2}{*}{$3.6$}  & \multirow{2}{*}{ }\\
				& & & & $0.00$ & $-1.06$ & $-0.59$ & & & &\\  \cmidrule{2-10}
				\multirow{2}{*}{$C_{10,\mu}^{\rm BSM}$} &  \multirow{2}{*}{$0.14$} & \multirow{2}{*}{$-0.23$} &  \multirow{2}{*}{$0.16$}  & $0.22$ & $0.36$  & $0.23$ & \multirow{2}{*}{$1.3$} &  \multirow{2}{*}{$1.2$} & \multirow{2}{*}{$1.6$} & \multirow{2}{*}{ }\\
				& & & & $0.06$ & $0.09$ & $0.09$ & & && \\  \cmidrule{2-10}
				\multirow{2}{*}{$C_{9,\mu}^{'\rm BSM}$} &  \multirow{2}{*}{$-0.01$} & \multirow{2}{*}{$0.20$} &  \multirow{2}{*}{$-0.03$}  & $0.07$ & $0.33$  & $0.05$ & \multirow{2}{*}{$0.9$} &  \multirow{2}{*}{$1.1$} & \multirow{2}{*}{$0.3$}  & \\
				& & & & $-0.09$ & $0.07$ & $-0.11$ & && & \\  \cmidrule{2-10}
				\multirow{2}{*}{$C_{10,\mu}^{'\rm BSM}$} &  \multirow{2}{*}{$-0.15$} & \multirow{2}{*}{$0$} &  \multirow{2}{*}{$0.03$}  & $-0.26$ & $0.08$  & $0.08$ & \multirow{2}{*}{$0.1$} &  \multirow{2}{*}{$0$} & \multirow{2}{*}{$0.3$} &  \\
				& & & & $-0.04$ & $-0.08$ & $-0.02$ & && & \\ 
				\toprule
				& \multicolumn{9}{c}{\textbf{New physics in the electron sector (Vector Axial basis)}} &    \\
				\cmidrule{2-10}
				& \multicolumn{3}{c}{Best-fit} & \multicolumn{3}{c}{1-$\sigma$ range} &
				\multicolumn{3}{c}{$\sqrt{\chi^2_{\rm SM} - \chi^2_{\rm best}}$}   &\\ \cmidrule{2-10}
				& \emph{HI}  & \emph{HS} & \emph{all}  & \emph{HI}  & \emph{HS} & \emph{all} & \emph{HI} & \emph{HS} & \emph{all} &   \\ \cmidrule{2-10}
				\multirow{2}{*}{$C_{9,e}^{\rm BSM}$} &  \multirow{2}{*}{$0.12$} & \multirow{2}{*}{$1.01$} &  \multirow{2}{*}{$0.15$}  & $0.24$ & $1.55$  & $0.26$ & \multirow{2}{*}{$0.7$} &  \multirow{2}{*}{$1.4$} & \multirow{2}{*}{$1.0$}  & \multirow{2}{*}{ }\\
				& & & & $0.00$ & $0.47$ & $0.04$ & & & &\\  \cmidrule{2-10}
				\multirow{2}{*}{$C_{10,e}^{\rm BSM}$} &  \multirow{2}{*}{$-0.09$} & \multirow{2}{*}{$-0.79$} &  \multirow{2}{*}{$-0.12$}  & $0.01$ & $-0.36$  & $-0.02$ & \multirow{2}{*}{$0.6$} &  \multirow{2}{*}{$1.2$} & \multirow{2}{*}{$0.9$} & \multirow{2}{*}{ }\\
				& & & & $-0.19$ & $-1.21$ & $-0.22$ & & && \\  \cmidrule{2-10}
				\multirow{2}{*}{$C_{9,e}^{'\rm BSM}$} &  \multirow{2}{*}{$0.15$} & \multirow{2}{*}{$0.20$} &  \multirow{2}{*}{$0.16$}  & $0.27$ & $0.33$  & $0.27$ & \multirow{2}{*}{$0.9$} &  \multirow{2}{*}{$1.4$} & \multirow{2}{*}{$1.0$}  & \\
				& & & & $0.03$ & $0.07$ & $0.05$ & && & \\  \cmidrule{2-10}
				\multirow{2}{*}{$C_{10,e}^{'\rm BSM}$} &  \multirow{2}{*}{$-0.14$} & \multirow{2}{*}{} &  \multirow{2}{*}{$-0.14$}  & $-0.04$ &   & $-0.40$ & \multirow{2}{*}{$0.9$} &  \multirow{2}{*}{} & \multirow{2}{*}{$1.0$} &  \\
				& & & & $-0.24$ &  & $-0.24$ & && & \\  \bottomrule
			\end{tabular}
		\end{center}
		\caption{\em Best fits turning on a single operator at a time in the vector-axial basis, using the `hadronic insensitive' observables `HI',  the `hadronic sensitive' observables `HS', or all the observables `all'.
		\label{tab:Fit2023bsaxial}}
	\end{table}%
Then, we proceed with a comprehensive fit involving multiple Wilson coefficients. Given the reduced sensitivity of our observables to electronic coefficients, we exclusively focus on the muonic operators.    
When turning on four BSM Wilson coefficients at once, the results of the global fit are:
	\beq
\begin{array}{rcl}
		C_ {b_L\mu _L}^{\text{BSM}} &=& -0.65 \pm 0.10 \ ,\\ 
		C_ {b_L\mu _R}^{\text{BSM}} &=& -0.93 \pm 0.09 \ ,\\ 
		C_ {b_R\mu _L}^{\text{BSM}} &=& 0.20 \pm 0.15 \ ,\\ 
		C_ {b_R\mu _R}^{\text{BSM}} &=& -0.16 \pm 0.28\ .\\ 
		\label{Globalmuonnew}
	\end{array}\quad \begin{aligned}
     &\chi^{2}_{SM}=264.99 \ ,\\
     &\tilde{\chi}^{2}=246.77\ .
     \end{aligned} \quad
	\rho = 
	\left(
\begin{array}{cccc}
 1 & 0.15 & -0.36 & -0.29 \\
 -0.15 & 1 & 0.28 & 0.39 \\
 -0.36 & 0.28 & 1 & 0.83 \\
 -0.29 & 0.39 & 0.83 & 1 \\
\end{array}
\right)
	.\eeq
We report the reduced chi-square:
\beq \label{eq:redglobalnew}
\frac{\chi^{2}_{SM}}{\text{\# d.o.f.}}=0.985 \ , \qquad \frac{\tilde{\chi}^{2}}{\text{\# d.o.f.}}=0.931 \ .
\eeq
When comparing the $\chi^2$ values, we find a deviation from the SM at the
4.3$\sigma$ level.  However, upon using the ``Pull" value, as defined in \cite{Blanke:2018yud}:
\begin{equation}
\label{eq:pullvalue}
   \text{Pull}= \sqrt{\text{CDF}^{-1}_{1}(\text{CDF}_{N_{par}}(\chi^2_{SM}-\tilde{\chi}^2))}\ ,
\end{equation}
we also take into account the number of parameters switched on. Here, $\text{CDF}_n$ stands for the cumulative distribution function of a $\chi^2$-distributed random variable with $n$ degrees
of freedom, and $N_{par}$ is the number of fitted
parameters. Using \eqref{eq:pullvalue}, the discrepancy with the SM is revealed at a slightly reduced significance level of $3.3 \, \sigma$.\\

 Interestingly, we observe that the results in \eqref{Globalmuonnew} agree within a $\sigma$  from the findings presented in \cite{DAlise:2022ypp}, when the global fit was performed already assuming  $R_K = R_{K^*} = 1$ long before the latest experimental values were available and before any other successive investigation. 

\subsection{Leptoquark motivated fit scenarios}
\label{sec:LQmotfit}
To conclude this section, we now analyse the possibility of NP scenarios involving the simultaneous activation of two Wilson coefficients. As we shall review in  \cref{sec:th}, LQ are primary examples of these type of models with the scalar $S_1=(\mathbf{\Bar{3}},\mathbf{1},1/3)$ and vector $U_1=(\mathbf{3},\mathbf{1},2/3)$ LQs being the relevant ones here because they turn on the following NP coefficients: 
\begin{align}
& S_1: C_ {b_L\mu_L}^{\text{BSM}},C_{b_L\mu_R}^{\text{BSM}}\ , \\
& U_1: C_{b_L\mu_L}^{\text{BSM}},C_{b_R\mu _R}^{\text{BSM}}\ .
\end{align}
The associated contour plots are given in \cref{fig:CLLLRRL}, corresponding to the fitted results:
\begin{align}
&S_1: \begin{cases} \begin{array}{rcl}
		C_ {b_L\mu _L}^{\text{BSM}} &=& -0.52 \pm 0.12 \ ,\\ 
		C_ {b_L\mu _R}^{\text{BSM}} &=& -0.64 \pm 0.14 \ ,
		\label{S1fit}
	\end{array}\end{cases}\quad {\rm{Pull}}= 3.3 \sigma \quad
	\rho = 
	\left(
\begin{array}{cc}
 1 & 0.54  \\
 0.54 & 1  \\
\end{array}
\right) \ , \\
&U_1: \begin{cases}
 \begin{array}{rcl}
		C_ {b_L\mu _L}^{\text{BSM}} &=& -0.31 \pm 0.09 \ ,\\ 
		C_ {b_R\mu _R}^{\text{BSM}} &=& 0.21 \pm 0.05 \ ,
		\label{U1fit}
	\end{array}
\end{cases}\quad {\rm{Pull}}= 1.9 \sigma
    \quad
	\rho = 
	\left(
\begin{array}{cc}
 1 & -0.47  \\
 -0.47 & 1  \\
\end{array}
\right) \ . 
\end{align} 
 We note that the $S_1$ model is favoured over the $U_1$ one when we consider both coefficients turned on simultaneously.  
\begin{figure}
\begin{subfigure}[h]{0.5\linewidth}
\includegraphics[height=6cm,keepaspectratio]{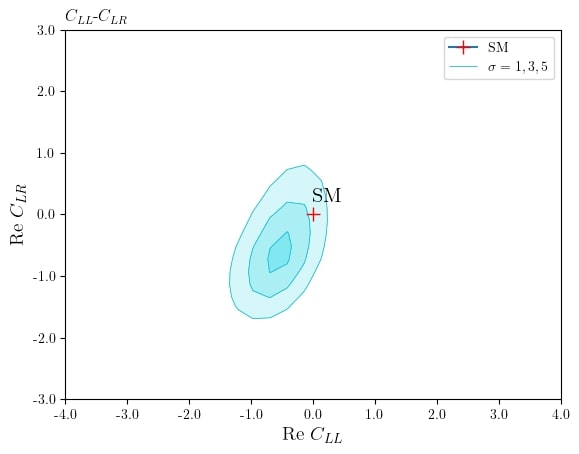}
\end{subfigure}
\hfill
\begin{subfigure}[h]{0.5\linewidth}
\includegraphics[height=6cm,keepaspectratio]{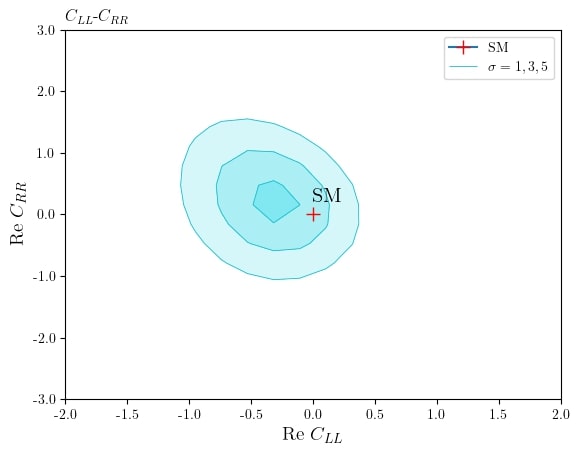}
\end{subfigure}%
\caption{Contour plots corresponding to the LQ motivated scenarios where we turn on 2 Wilson coefficients at a time, while fixing the remaining coefficients to zero. We use the full list of observables given in Appendix \ref{appbtoc} and present the contour plots corresponding to the $1$, $3$, and $5$-$\sigma$ confidence levels, along with the corresponding SM value.}
\label{fig:CLLLRRL}
\end{figure} 
 
	\section{\texorpdfstring{$b \to c$ } ~ Flavour sector}\label{btoc}
	
So far, we concentrated on Flavor Changing Neutral Currents (FCNC) process \(b \to s \ell^+ \ell^-\), it is therefore time to tackle the Flavor Changing Charged Currents (FCCC) sector via the decays \(b \to c \ell^- \nu_{\ell}\). For these, the most general low-energy effective Hamiltonian including operators up to dimension six is:
\beq
\label{WETRDRDS}
   \mathscr{H}_{\rm{eff}}=-V_{cb}\frac{\alpha_{\rm em}}{4 \pi v^2}[ (1+C_{V_{L}}) \mathcal{O}_{V_L}+C_{V_{R}} \mathcal{O}_{V_R}+C_{S_{L}} \mathcal{O}_{S_L}+C_{S_{R}} \mathcal{O}_{S_R}+C_{T} \mathcal{O}_{T} + \rm{h.c.} ] \ ,
\eeq
with
\begin{equation}
   \begin{split}
     & \mathcal{O}_{V_L}^{\ell} = (\bar{c}_L \gamma^\mu b_L)(\bar{\ell}_L \gamma_\mu \nu_{\ell L}) \ ,  \\
      &\mathcal{O}_{V_R}^{\ell} = (\bar{c}_R \gamma^\mu b_R)(\bar{\ell}_L \gamma_\mu \nu_{\ell L}) \ , \\
      &\mathcal{O}_{S_L}^{\ell} = (\bar{c}_L b_R)(\bar{\ell}_R \nu_{\ell L}) \ , \\
      &\mathcal{O}_{S_R}^{\ell} = (\bar{c}_R b_L)(\bar{\ell}_R \nu_{\ell L}) \ , \\
      &\mathcal{O}_T^{\ell} = (\bar{c}_R \sigma^{\mu\nu} b_L)(\bar{\ell}_R \sigma_{\mu\nu} \nu_{\ell L}) \ .
   \end{split}
   \label{eq:operators}
\end{equation}  
The NP contributions are, differently from the FCNC case, are directly encoded in the Wilson coefficients (WCs), denoted as \(C_X\).  We now turn to contributions stemming from the tau sector only. 

\subsection{Current status of the \texorpdfstring{$b\to c$}{b to c} Observables}

The semi-tauonic decays of $B$ mesons, specifically $B \to D^{(\ast)} \tau \overline{\nu}$, are crucial processes for probing lepton flavor universality violation (LFUV), complementing the investigation of $B \to K^{(\ast)} \ell \ell$ discussed in the preceding section. The ratios
\begin{align}
  R_D &= \frac{\text{BR}( B\rightarrow D \,\tau\, \overline{\nu}_\tau)}{\text{BR}(B\rightarrow D\, \ell\,\overline\nu_\ell)}, \\
  R_{D^{\ast}} &= \frac{\text{BR}(B\rightarrow D^{\ast} \tau \,\overline{\nu}_\tau)}{\text{BR}(B\rightarrow D^{\ast} \ell\,\overline{\nu}_\ell)}, \quad \ell= e, \mu,
\end{align}
have been computed with high precision. This precision is achieved through the remarkable suppression of  hadronic uncertainties associated with the strong interaction in the $B \to D^{(\ast)}$ transitions.

The experimental measurements, as detailed in \cref{tab:RD_exps}, consistently reveal deviations from Standard Model predictions. Notably, the characteristic pattern is that the experimental values of $R_{D^{(\ast)}}$ exceed the SM predictions, implying a systematic violation of lepton flavor universality (LFU). In \Cref{tab:RD_exps}, we provide a concise overview of the current status of independent measurements of $R_{D}$ and $R_{D^{\ast}}$ conducted by various collaborations. We also include the latest result on $R_{D^*}$ from Belle II from the recent presentation in \cite{belle2talk}. \\ 
In addition to the ratios $R_D$ and $R_{D^{\ast}}$, significant constraints arise from the $B_c$ meson branching ratio $\text{BR}(B_c \rightarrow \tau \nu_\tau)$. The upper bound on the branching ratio $\text{BR}(B_c \rightarrow \tau \nu_\tau)$ is given as $\text{BR}(B_c \rightarrow \tau \nu_\tau) \leq 0.3$ according to \cite{Alonso:2016oyd}. Alternatively, a more conservative limit of $0.6$ is suggested in \cite{Blanke:2018yud}, while a more stringent bound of $0.1$, based on LEP data, is presented in \cite{Akeroyd:2017mhr}.

In our analysis, we also consider the measurement of the $D^{\ast}$ meson longitudinal polarisation fraction, denoted as $F_L^{D^{\ast}}$, with a value of $0.60 \, \pm 0.08 \, \pm 0.04$, as reported by the Belle collaboration \cite{Belle:2019ewo}. This result is in agreement within approximately $1.7\sigma$ with the SM prediction.

\begin{table}[t]
\centering
\newcommand{\bhline}[1]{\noalign{\hrule height #1}}
\renewcommand{\arraystretch}{1.5}
\addtolength{\tabcolsep}{5pt} 
   \scalebox{1}{
  \begin{tabular}{lccc} 
  \bhline{1 pt}
  \rowcolor{white}
 Experiment  &$R_{D^\ast}$ & $R_{D}$ & Correlation  \\  \hline 
BaBar  \cite{BaBar:2012obs,BaBar:2013mob}  & $0.332\pm 0.024\pm 0.018$ & $0.440\pm 0.058 \pm 0.042$ & $-0.27$ \\
\hline
Belle \cite{Belle:2015qfa} & $0.293\pm 0.038\pm 0.015$ & $0.375\pm 0.064\pm 0.026$ & $-0.49$\\
\hline
Belle \cite{Belle:2016dyj,Belle:2017ilt}
& $0.270\pm 0.035^{+0.028}_{-0.025}$ & --& -- \\
\hline
Belle \cite{Belle:2019gij,Belle:2019rba}
& $0.283\pm 0.018\pm 0.014$ & $0.307\pm 0.037\pm 0.016$ & $-0.51$\\
\hline
LHCb\cite{LHCb:2023zxo} & $0.281 \pm 0.018 \pm 0.024$ &$0.441\pm 0.060 \pm 0.066$ & $-0.43$ \\ 
\hline
LHCb\cite{LHCb:2023uiv} & $0.257 \pm 0.012 \pm 0.018$ &-- & -- \\ 
\hline
Belle II \cite{belle2talk} & $0.267^{+0.041}_{-0.039} {}^{+0.028}_{-0.033}$ &-- & -- \\ 
\hline
 World average \cite{HeavyFlavorAveragingGroup:2022wzx} & $0.284\pm 0.012$ &$0.357\pm 0.029$ & $-0.37$ \\ 
\bhline{1 pt}
   \end{tabular}
}   
\addtolength{\tabcolsep}{-5pt} 
 \caption{ \label{tab:RD_exps}
 Current status of the independent experimental $R_{D^{(\ast)}}$ measurements. 
 The first and second errors are statistical and systematic, respectively. 
 }
\end{table}

\subsection{Analysis of the \texorpdfstring{$b \to c$}{b to c} Observables}

Using the effective Hamiltonian from Eq. \eqref{WETRDRDS}, where the NP Wilson coefficients are defined at the renormalisation scale $\mu=\mu_b=4.18 \,{\rm GeV}$, we employ the updated numerical formulas provided in \cite{Iguro:2022yzr}:
\begin{align}
\label{RDform}
    \frac{R_{D}}{R_{D}^{{SM}}}&= \abs{1+C_{V_L}+C_{V_R}}^2+1.01 \abs{C_{S_R}+C_{S_L}}^2+0.84 \abs{C_{T}}^2\nonumber\\&+1.49 Re[(1+C_{V_L})(C_{S_R}^\ast+C_{S_L}^\ast)]+1.08 Re[(1+C_{V_L}+C_{V_R})C_{T}^\ast] \ ,
\end{align}
\begin{align}
\label{RDsform}
    \frac{R_{D^\ast}}{R^{{SM}}_{D^\ast}}&= \abs{1+C_{V_L}}^2+\abs{C_{V_R}}^2+0.04\abs{C_{S_L}-C_{S_R}}^2+16.0\abs{C_T}^2\nonumber\\&-0.11 Re\left[\left(1+C_{V_L}-C_{V_R}\right)\left(C^\ast_{S_L}-C^\ast_{S_R}\right)\right] -1.83 Re \left[(1+C_{V_L})C^*_{V_R} \right]\nonumber\\&-5.17 Re\left[\left(1+C_{V_L}\right)C^\ast_T \right]+6.60 Re \left[ C_{V_R}C_T^* \right ] \ ,
\end{align}
\begin{align}
\label{FLsform}
    \frac{F_L^{D^\ast}}{F_{L , SM}^{D^\ast}}&= \Big(\frac{R_{D^\ast}}{R^{{SM}}_{D^\ast}} \Big)^{-1}\Big(\abs{1+C_{V_L}-C_{V_R}}^2+0.08\abs{C_{S_L}-C_{S_R}}^2+6.90\abs{C_T}^2\nonumber\\&-0.25 Re\left[\left(1+C_{V_L}-C_{V_R}\right)\left(C^\ast_{S_L}-C^\ast_{S_R}\right)\right] -4.30 Re\left[\left(1+C_{V_L}-C_{V_R}\right)C^\ast_T \right]\Big)\ ,
\end{align}
\begin{equation}
    \begin{aligned}
        \frac{\text{BR}(B_c^+\rightarrow \tau^+\nu_\tau)}{\text{BR}(B_c^+\rightarrow \tau^+\nu_\tau)_{SM}}= \abs{1+C_{V_L}-4.35(C_{S_L}-C_{S_R})}^2 \ .
       \label{Bctaunu}
    \end{aligned}
    \end{equation}
The SM values that we used for these quantities are given by \cite{Iguro:2022yzr,HeavyFlavorAveragingGroup:2022wzx}
\begin{equation}
\begin{aligned}
    R_D^{SM}=0.298 \ \ , \quad
    R^{{SM}}_{D^\ast}=0.254 \ ,\quad
    F_L^{D^*}=0.464 \ ,\quad
    \text{BR}(B_c^+\rightarrow \tau^+\nu_\tau)_{SM}=0.022 \ .
\label{SMvalues}
\end{aligned}
\end{equation}
Using the expressions for $R_D$ and $R_{D^*}$, we illustrate in Fig. \ref{fig:RDRDs1coeff} the parametric curves corresponding to the single NP Wilson coefficient in the $R_{D}-R_{D^*}$ plane, along with the current experimental world average reported in \cref{tab:RD_exps}.

Using the experimental values for the observables in Eqs. (\ref{RDform}--\ref{Bctaunu}) and considering the 2-$\sigma$ interval, we establish constraints on the Wilson coefficients when turning on a single coefficient at a time. We also explore scenarios motivated by LQ models, defined by the relations $C_{S_L}(\Lambda_{LQ})=+ 4 C_T(\Lambda_{LQ})$ and $C_{S_L}(\Lambda_{LQ})=- 4 C_T(\Lambda_{LQ})$, at the LQ scale $\Lambda_{LQ} \approx 2 \, {\rm TeV}$. After accounting for the renormalisation group (RG) running from $\Lambda_{LQ}$ to $\mu_b$, these relations translate into $C_{S_L}(\mu_{b})=8.4 C_T(\mu_b)$ and $C_{S_L}(\mu_{b})=-8.9 C_T(\mu_b)$ respectively \cite{Iguro:2022yzr,Capdevila:2023yhq}. The constraints at $\mu_b$ are:
\begin{align}
\label{Constraintbtoc1}
    &C_{V_L} \in [0.01,0.10] \ , \\
    \label{Constraintbtoc2}
    &C_{S_R} \in [0.20,0.23] \ , \\ 
    \label{Constraintbtoc3}
    &C_{S_L}=-8.9C_{T} \in [0.05,0.24] \ .
\end{align}
As shown in \cref{fig:RDRDs1coeff}, the curve corresponding to $C_{S_L}(\Lambda_{LQ})=+4 C_T(\Lambda_{LQ})$ fails to explain the experimental data at the $2-\sigma$ level. It is worth mentioning that, even when considering both $C_{S_L}$ and $C_T$ individually, we are unable to reconcile the data. Therefore, we refrained from including the corresponding parametric curves explicitly in the figure.
\\It is also relevant to consider the scenario where purely imaginary values are allowed for the LQ-motivated coefficients. In this case, we find:
\begin{align}
\label{Constraintbtoc4}
    &|\rm{Im}C_{S_L}|=8.9|\rm{Im}C_{T}| \in [0.31,0.62] \ , \\
    \label{Constraintbtoc5}
    &|\rm{Im}C_{S_L}|=8.4|\rm{Im}C_{T}| \in [0.30,0.62] \ .
\end{align}
\begin{figure}[h!]
    \centering
\includegraphics[scale=0.6]{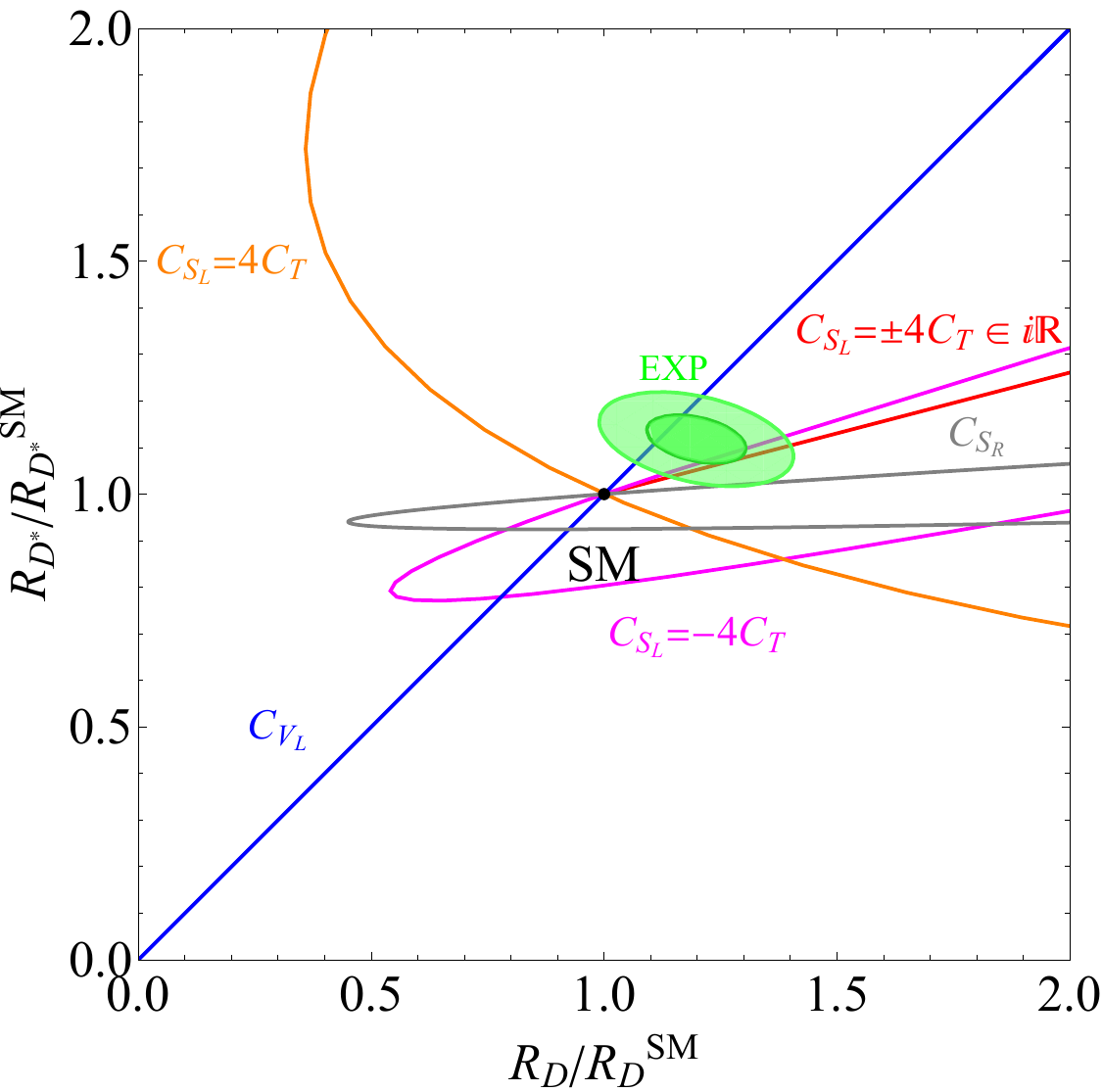}
    \caption{Quantitative behaviour of $R_{D}$ vs $R_D^*$, normalised by their SM value, obtained by switching on one NP coefficient at a time. The experimental point is depicted with $1\sigma$ (light green) and $2\sigma$ (dark green) contours. The condition $C_{S_L}= \pm4 C_T$ are imposed at $2 \,{\rm TeV}$ and we provide the parametic curves after accounting for the RG running to the low energy scale $\mu_b$.} 
    \label{fig:RDRDs1coeff}
\end{figure}
 \newpage
\subsection{Fit Analysis}
We employ the {\sc{Flavio}} package \cite{Straub:2018kue} to perform a fitting analysis of the Wilson coefficients at the scale $\mu_b$ to match the experimental data.\\
In order to see how NP scenarios improve the fit, especially when turning on more than one coefficient at a time, we use the ``Pull” value defined in Eq. \eqref{eq:pullvalue}, which is expressed in terms of standard deviations $\sigma$ \cite{Blanke:2018yud}.
Similar to our approach in the $b \to s$ sector, our initial analysis focuses solely on the observables defined in Eqs. (\ref{RDform}--\ref{Bctaunu}). Also in this case we refer to these observables as ``hadronic insensitive" as they are characterised by a strong suppression of hadronic uncertainties and thus predicted in the SM with an error up to few percent. Subsequently, we extend our analysis to include the effects of all available data summarised in Tables (\ref{tab:bcbr_obs} -- \ref{tab:bcbr_tot}).\\
We systematically turn on each Wilson coefficient individually, assuming real values. Our results in \cref{tab:btocfit} align with those reported in \cite{Iguro:2022yzr} within a 1-$\sigma$ level.
We notice a preference for NP coming from a nonzero value of the $C_{V_L}$  Wilson coefficient  with a significance of $4.3 \sigma$.
\begin{table}[pt]
    \begin{center}
        \begin{tabular}{cccccrc}
            \toprule
            & \multicolumn{6}{c}{\textbf{New physics in the tau sector }} \\
            \cmidrule{2-7}
            & \multicolumn{2}{c}{Best-fit} & \multicolumn{2}{c}{1-$\sigma$ range} &
            \multicolumn{2}{l}{$\sqrt{\chi^2_{\rm SM} - \chi^2_{\rm best}}$} \\
            \cmidrule{2-7}
            & \emph{HI} & \emph{all} & \emph{HI} & \emph{all} & \emph{HI} & \emph{all} \\
            \cmidrule{2-7}
            \multirow{2}{*}{$C_{V_L} $} & \multirow{2}{*}{$0.08$} & \multirow{2}{*}{$0.08$} & $0.09$ & $0.09$ & \multirow{2}{*}{$4.3$} & \multirow{2}{*}{$4.8$} \\
            & & & $0.07$ & $0.07$ & & \\ \cmidrule{2-7}
            \multirow{2}{*}{$C_{V_R} $} & \multirow{2}{*}{$-0.06$} & \multirow{2}{*}{$-0.07$} & $-0.04$ & $-0.05$ & \multirow{2}{*}{$2.0$} & \multirow{2}{*}{$2.7$} \\
            & & & $-0.08$ & $-0.09$ & & \\ \cmidrule{2-7}
            \multirow{2}{*}{$C_{S_L} $} & \multirow{2}{*}{$0.16$} & \multirow{2}{*}{$0.16$} & $0.20$ & $0.20$ & \multirow{2}{*}{$2.6$} & \multirow{2}{*}{$2.8$} \\
            & & & $0.12$ & $0.12$ & & \\ \cmidrule{2-7}
            \multirow{2}{*}{$C_{S_R} $} & \multirow{2}{*}{$0.19$} & \multirow{2}{*}{$0.20$} & $0.22$ & $0.23$ & \multirow{2}{*}{$3.8$} & \multirow{2}{*}{$4.1$} \\
            & & & $0.16$ & $0.17$ & & \\
            \cmidrule{2-7}
            \multirow{2}{*}{$C_{T} $} & \multirow{2}{*}{$-0.03$} & \multirow{2}{*}{$-0.03$} & $0.03$ & $0.02$ & \multirow{2}{*}{$3.5$} & \multirow{2}{*}{$4.0$} \\
            & & & $-0.09$ & $-0.08$ & & \\
            \bottomrule
        \end{tabular}
    \end{center}
    \caption{\em Best fits at the scale $\mu_b$ turning on a single operator at a time using the `hadronic insensitive' observables `HI' or all the observables `all'. In analogy with the analysis of $b \to s$ data, here with HI observables we refer to Eqs. (\ref{RDform}--\ref{Bctaunu}).  The full list of observables can be found in Appendix~\ref{appbtoc}.
        \label{tab:btocfit}}
\end{table}
\\
Upon allowing the Wilson coefficients to take on complex values, a significant improvement is observed in the fitting for $C_{V_R}$ and $C_{S_L}$, while no enhancement is obtained for $C_{V_L}$, $C_{S_R}$, and $C_{T}$. When considering the set of observables given in Eqs. (\ref{RDform}--\ref{Bctaunu}), we get
\begin{align}
    \label{imagCSLCT}
    &C_{V_R}=0.01 \pm 0.43 i  \pm(0.02+0.04 i) \, \rightarrow 3.9 \sigma  \ , \\
   & C_{S_L}=-0.12 \pm 0.63 i  \pm(0.08+0.07 i) \, \rightarrow 3.3 \sigma \ .
\end{align}
Slightly higher Pull values are achieved when considering the complete set of observables.\\
\subsection{LQ motivated fit scenarios}
We now proceed to present the results of the fit analysis, considering additional scenarios motivated by the LQ models that we will explore in the next section. We  focus exclusively on the set of HI observables specified in Eqs. (\ref{RDform}--\ref{Bctaunu}). \\
The LQ models along with the combinations of Wilson coefficients not previously considered, evaluated at the LQ scale $\Lambda_{LQ}$, are:
\begin{align}
    \label{LQWCs}
    &S_1: C_{V_L}, \, C_{S_L} = -4C_T \ , \\
    &U_1: C_{V_L}, \, C_{S_R} \ ,\\
    &R_2:  C_{S_L} = 4C_T \ .
\end{align}
Here we have the following SM quantum numbers for $R_2=(\mathbf{3},\mathbf{2},7/6)$. 
After accounting for the RG running from $\Lambda_{LQ}$ to $\mu_b$, the relations $C_{S_L}(\Lambda_{LQ}) = +4C_T(\Lambda_{LQ}) $ and $C_{S_L} (\Lambda_{LQ}) = -4C_T(\Lambda_{LQ}) $ become respectively $C_{S_L}(\mu_{b})=8.4 C_T(\mu_b)$ and $C_{S_L}(\mu_{b})=-8.9 C_T(\mu_b)$ \cite{Iguro:2022yzr,Capdevila:2023yhq}. 
The best fit for these two scenarios gives 
\begin{align}
    \label{fitCST}
    & C_{S_L}(\mu_{b})=8.4 C_T(\mu_b)=-0.01  \pm 0.04  \, \rightarrow \ 0.2 \sigma \ , \\
       & C_{S_L}(\mu_{b})=8.4 C_T(\mu_b)=\pm 0.53 i \pm 0.05 i \, \rightarrow \ 4.2 \sigma \ , \\
   & C_{S_L}(\mu_{b})=8.4 C_T(\mu_b)=-0.07\pm 0.55 i \pm(0.04 + 0.05 i)  \, \rightarrow \, 4.0\sigma \ , \\
    & C_{S_L}(\mu_{b})=-8.9 C_T(\mu_b)=0.18 \pm 0.03  \, \rightarrow \ 4.0 \sigma \ , \\
    & C_{S_L}(\mu_{b})=-8.9 C_T(\mu_b)=\pm 0.53 i \pm 0.05 i \, \rightarrow \ 4.1 \sigma \ , \\
    & C_{S_L}(\mu_{b})=-8.9 C_T(\mu_b)=0.07\mp 0.44 i \pm(0.08 + 0.13 i)  \, \rightarrow \, 3.8\sigma \ .
\end{align}
In the first case, the fit indicates a clear preference for complex values, while in the second scenario, it does not. However, in general, the LQs $S_1$ and $U_1$ turn on two independent Wilson coefficients simultaneously. While the coefficients $C_{V_L}$ and $C_{S_R}$ are preferred to be real, we consider the two cases when the combination $C_{S_L}=-8.9C_T$ is either real or purely imaginary. We show the contour plots in \cref{fig:CSLCT} and \cref{fig:CSRCVL} corresponding to the fit results:
 \begin{align}
&S_1: \begin{cases}
 \begin{array}{rcl}
		C_ {V_L} &=& 0.09 \pm 0.04 \ ,\\ 
		C_ {S_L} &=& -8.9 C_T= -0.01 \pm 0.09 \ ,
		\label{VLSLreal}
	\end{array}
\end{cases}\; {\rm{Pull}}= 3.9 \sigma \quad
	\rho = 
	\left(
\begin{array}{cc}
 1 & -0.93  \\
 -0.93 & 1  \\
\end{array}
\right) \ 
	,\\
 & S_1: \begin{cases}
 \begin{array}{rcl}
		C_ {V_L} &=& 0.06 \pm 0.03 \ ,\\ 
		C_ {S_L} &=& -8.9 C_T= (\mp0.34 \pm 0.13)i \ ,
		\label{VLSLimag}
	\end{array}
\end{cases}\; {\rm{Pull}}= 4.0 \sigma \quad 
	\rho = 
	\left(
\begin{array}{cc}
 1 & 0.84  \\
 0.84 & 1  \\
\end{array}
\right) \ ,\\
 & U_1: \begin{cases}
 \begin{array}{rcl}
		C_{V_L} &=& 0.07 \pm 0.02 \ ,\\ 
		C_{S_R} &=& 0.06 \pm 0.06 \ ,
		\label{VLSR}
	\end{array}
\end{cases}\; {\rm{Pull}}= 4.0 \sigma    \quad
	\rho = 
	\left(
\begin{array}{cc}
 1 & -0.77  \\
 -0.77 & 1  \\
\end{array}
\right)
	\ . \end{align}
\begin{figure}
\begin{subfigure}[h]{0.5\linewidth}
\includegraphics[scale=0.58]{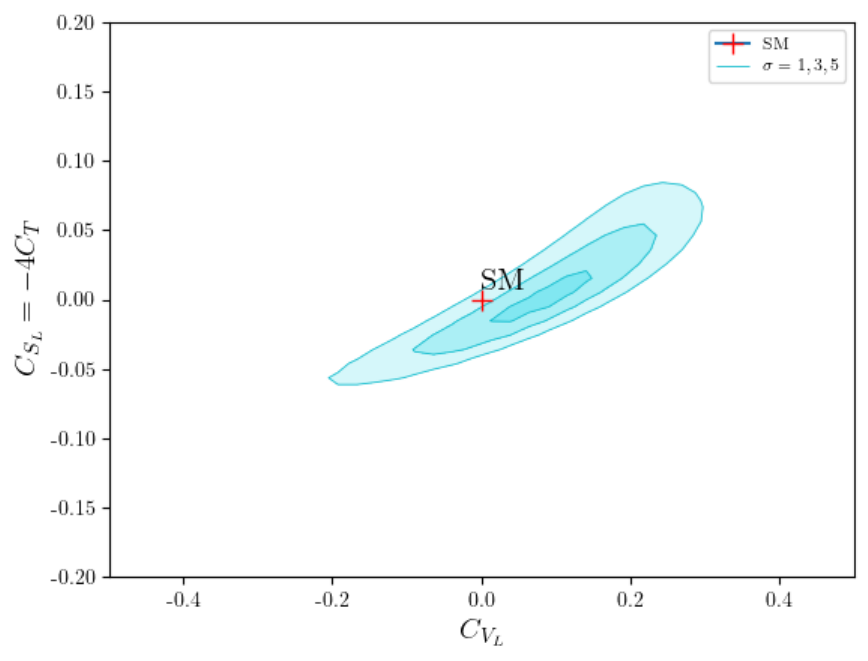}
\end{subfigure}
\hfill
\begin{subfigure}[h]{0.5\linewidth}
\includegraphics[scale=0.58]{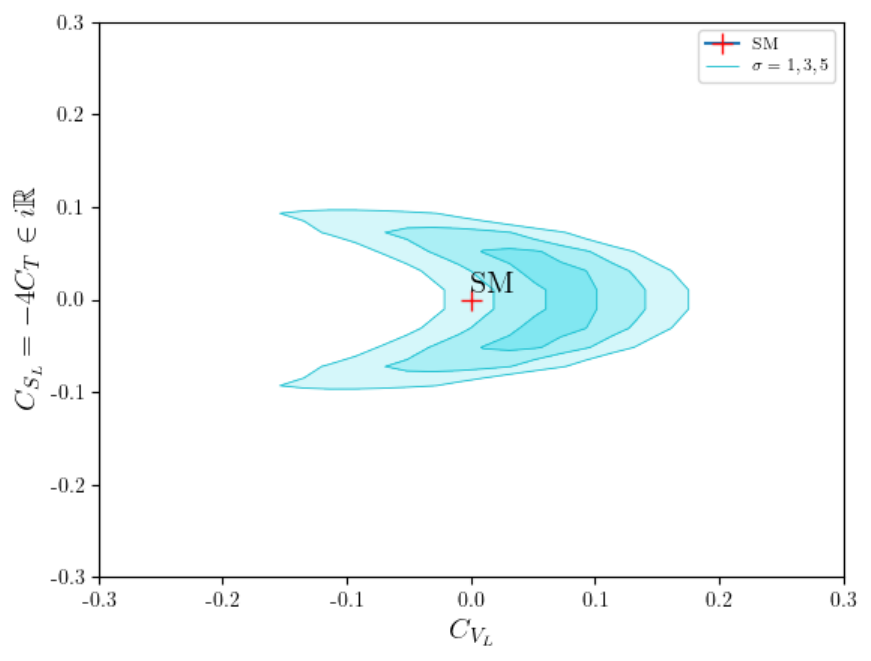}
\end{subfigure}%
\caption{Contour plots illustrating the behavior of NP coefficients $C_{V_L}$, $C_{S_L}$, and $C_T$, considering the theoretical relation $C_{S_L}(\Lambda_{LQ})=-4C_T(\Lambda_{LQ})$ predicted by the LQ $S_1$. The left panel represents the case where $C_{S_L}=-4C_T$ is real, while in the right panel it is purely imaginary. We present contour plots corresponding to the $1$, $3$, and $5$-$\sigma$ confidence levels, along with the corresponding SM value.}
\label{fig:CSLCT}
\end{figure}
\begin{figure}[h]
    \centering
    \includegraphics[scale=0.6]{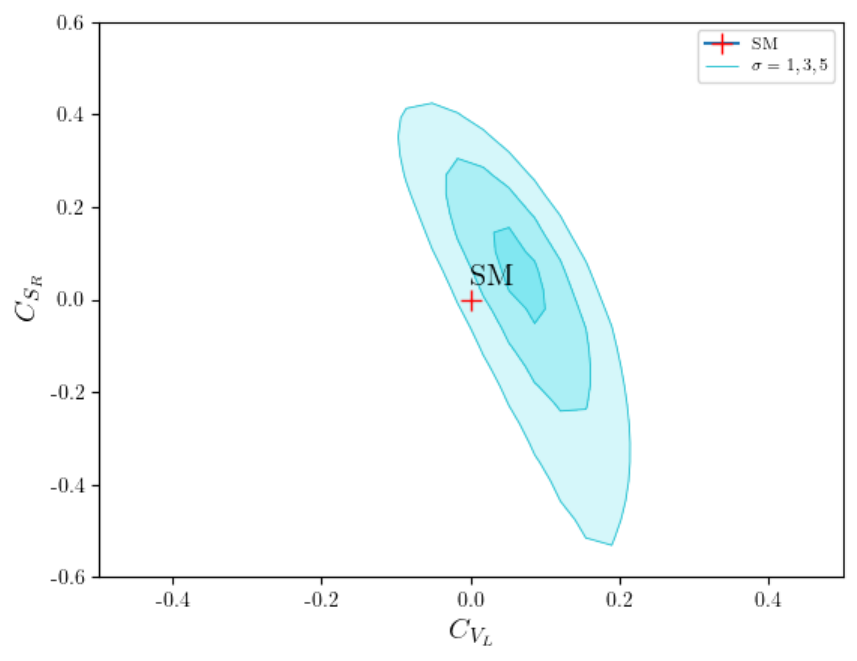}
    \caption{Contour plot for NP coefficients $C_{V_L}$ and $C_{S_R}$. We present contour plots corresponding to the $1$, $3$, and $5$-$\sigma$ confidence levels, along with the corresponding SM value.}
    \label{fig:CSRCVL}
\end{figure}

Differently from the case of $b\rightarrow s$ the HI observables hint to the presence of new physics in the  $b\rightarrow c$ sector at four sigma level, provided future experimental analyses will avail the current results.

\FloatBarrier

	\section{The New Physics Landscape}\label{sec:th}
 The burning question for present and future experiments in particle physics is: {\it Where is the scale of new physics?}  This  overarching goal of the present section is to answer this question by  comparing time-honored extensions of the SM, such as $Z^{\prime}$ and LQs models, with  $b\to s$  and $b\to c$ observables.


 \subsection{Shifting up the \texorpdfstring{$Z^\prime$} - boson}
	$Z^\prime$ bosons have a long history \cite{1310.1082,1311.6729,1403.1269,1501.00993,1503.03477,1503.03865,1505.03079,1506.01705,1508.07009,1509.01249,1510.07658,1512.08500,1601.07328,1602.00881,1604.03088,1608.01349,1608.02362,1611.03507,1702.08666,1703.06019} and can affect FCNC, with impact for the ratios $R_{K^{(*)}}$, via their interactions   
		\beq [g_{bs} (\bar s \gamma_\mu P_L b) +\hbox{h.c.}] + g_{\mu_L} (\bar\mu \gamma_\mu P_L \mu) \ ,\eeq
 which generates the $b \to s \ell^+ \ell^-$ transitions. At tree level, a $Z^\prime$ with the above interactions and mass $M_{Z^\prime}$ yields
 \begin{equation}
     C^{\rm{BSM}}_{b_L \mu_L} =-\frac{4 \pi v^2}{V_{tb}V^\ast_{ts}\alpha_{\rm em}} \frac{ g_{bs} g_{\mu_L}}{M_{Z'}^2} \ .
 \end{equation}  
With the current measurements, this scenario pushes the lower bound for the $Z'$ mass to about $50$ TeV, assuming the product of the couplings to be of order unity for perturbative realizations.  For this estimate we used the analytic results in \eqref{Constraintbtos1} stemming from a direct comparison with the HI observables. We checked that if we consider the global fit with including the HS observables the result does not change.

	\subsection{Shrinking the Leptoquarks landscape }\label{LQ}

	Leptoquarks as bosonic particles carrying both quark and lepton numbers,  have been extensively discussed in the literature ~\cite{ Gripaios:2009dq,Gripaios:2014tna,deMedeirosVarzielas:2015yxm, Crivellin:2017zlb, Hiller:2014yaa,Sahoo:2015wya, Alonso:2015sja,Barbieri:2015yvd, Becirevic:2016yqi, Cai:2017wry,Sahoo:2016pet,Barbieri:2016las,  Buttazzo:2017ixm, Marzocca:2018wcf,Becirevic:2018afm,Vignaroli:2018lpq,Vignaroli:2019lkg,BhupalDev:2020zcy,Crivellin:2022mff} as compelling model building especially when trying to explain possible flavor anomalies. Here we list the relevant LQs, classified by their quantum numbers under the SM gauge group $(SU(3)_c, SU(2)_L, U(1)_Y)$, and indicate for each LQ its compatibility with the up-to-date  flavor data.

\subsubsection{Scalar and Vector Leptoquarks}

 We consider three types of scalar LQs ($S_1$, $R_2$, $S_3$) and three types of vector LQs ($U_1$, $V_2$, $U_3$) which may still be compatible with the new picture emerging from the flavor data.  We enforce baryon number conservation  to prevent proton decay and further ignore couplings with right-handed neutrinos.  

\begin{itemize}

\item{$S_1=(\mathbf{\Bar{3}},\mathbf{1},1/3)$}\\
The interaction of the weak singlet scalar LQ reads:
\begin{equation}
\begin{split}
    &y^{LL}_{1\,ij}\Bar{Q}^{C\ i,a}_L S_1 (i\tau^2)^{ab}L^{j,b}_L+\ y^{RR}_{1\,ij}\Bar{u}^{C\ i}_R S_1 e^j_R +{\rm{h.c.}}\ .
    \end{split}
\end{equation}

 The $S_1$ interactions can mediate at tree-level charged transitions $b \to c \tau \bar \nu_{\tau}$.
, with effective couplings:
\begin{align}
\begin{split}
   & C_{S_L} = -4 C_T=-\frac{v^2}{2 V_{cb}}\frac{y_{1\,b\tau}^{LL} (y_{1\,c\tau}^{RR\,*})}{m^2_{S_1}} \ ,\\
    & C_{V_L} = \frac{v^2}{2 V_{cb}}\frac{y_{1\,b\tau}^{LL} (V y_{1}^{LL\,*})_{c\tau}}{m^2_{S_1}} \ .
    \end{split}
\end{align}
With these relations, $S_1$ can explain the observed anomalies in $R_{D^{(*)}}$.
Neutral current transitions $b \to s \mu \mu$ can be also generated, but only radiatively:
  \begin{align}
C^{\text{BSM}} _{b_L \mu_R}&=\frac{m^2_t}{8\pi \alpha_{\rm em} m^{2}_{S_{1}}}\left( V^*y^{LL}_{1}\right)_{t \mu} \left( V^*y^{LL}_{1}\right)^*_{t \mu}-\frac{v^2}{16\pi \alpha_{\rm em} m^{2}_{S_{1}} } \frac{\left( y^{LL}_{1}\cdot y^{\dagger\, LL}_{1}\right)_{b s}}{V_{tb}V^\ast_{ts}} \left( y^{\dagger\, LL}_{1} \cdot y^{LL}_{1}\right)_{\mu \mu} \ , \\
C^{\text{BSM}} _{b_L \mu_L}&= \frac{m^2_t}{8\pi \alpha_{\rm em} m^{2}_{S_{1}}} \left( y^{RR}_{1\,t \mu}\right) \left( y^{*\, RR}_{1\, t \mu }\right) \left[\log\left( \frac{m^{2}_{S_{1}}}{m^{2}_{t}}\right)-f(x_t) \right]-\\&- \frac{v^2}{16\pi \alpha_{\rm em} m^{2}_{S_{1}} } \frac{\left( y^{LL}_{1}\cdot y^{\dagger\, LL}_{1}\right)_{b s}}{V_{tb}V^\ast_{ts}} \left( y^{\dagger\, LL}_{1} \cdot y^{LL}_{1}\right)_{\mu \mu}  \ ,
\end{align}
where 
$$
f(x_t)=1+\frac{3}{x_t-1}\left(  \frac{\log(x_t)}{x_t-1}-1\right), \quad x_t=\frac{m^2_t}{M^2_W} \ .
$$
Therefore with order one couplings, we find that $S_1$ can explain both the $b \to c$ anomalies and the SM-like results for the $b \to s$ processes. The associated mass scale are of the order of a few  $\rm{TeV}$: 
\begin{equation}
   1.5 \, {\rm{TeV}} \lesssim m_{S_1}\lesssim   3.6 \, \rm{TeV} \ .
\end{equation}
We used the analytic results for the HI observables for both type of transitions. With the upper bound coming from explaining the presence of $b\to c$ anomalies and the lower bound from not upsetting the SM-like $b \to s$ processes. If, in the future, all anomalies would disappear the upper bound will be lifted.

\item{$R_2=(\mathbf{3},\mathbf{2},7/6)$}\\
$R_2$ is a weak doublet with interactions preserving the baryon number that are:
\begin{equation}
    -y^{RL}_{2\,ij}\Bar{u}^i_R R^a_2(i\tau^2)^{ab}L^{j,b}_L+\ y^{LR}_{2\, ij}\Bar{e}^i_R R^{a\ \ast}_2\  Q^{j,a}_L+{\rm{h.c.}}\ .
\end{equation}
This model can accommodate the data on $R_{D^{(*)}}$  by generating at tree level the charged transition $b \to c \tau \bar{\nu_{\tau}}$, with an effective coefficient:
\begin{equation}
    C_{S_L} = 4C_T= \frac{ v^2}{2 V_{cb}}\frac{y_{2\,c\tau}^{LR} (y_{2\,b\tau}^{RL})^*}{m^2_{R_2}} \ .
\end{equation}
In fact one can generate, still at tree level,  the coefficient
\begin{equation}
  C^{\text{BSM}} _{b_L \mu_R}=-\frac{2 \pi v^2}{V_{tb}V^\ast_{ts}\alpha_{\rm em}}\frac{y^{LR}_{2\,s\mu} y^{LR\ \ast}_{2\,b\mu}}{m^2_{R_2}} \, ,
\end{equation} 
relevant for the $b \to s$ processes.
As a result this LQ can accommodate  the $b\to c$ anomalies while agreeing with the $b\to s$ results only for specific structures of couplings such as small couplings to muons than to taus. 
    If, however, we assume all the couplings of order one, the $b \to s$ data would require over $20~$TeV for the mass of $R_2$ which would mean that the $b \to c$ data are not compatible with this scenario. 

\item{$S_3=(\mathbf{\Bar{3}},\mathbf{3},1/3)$}\\
The Lagrangian of the weak triplet is given by:
\begin{equation}
    y^{L}_{3\,ij}\Bar{Q}^{C\ i,a}_L\ (i\tau^2)^{ab}\left(\tau^k S^k_3\right)^{bc}\ L^{j,c}_L+ {\rm{h.c.}}
\end{equation}
Note that the couplings to diquarks are suppressed in order to guarantee   proton stability. 
$S_3$ can mediate neutral current transitions $b \to s \mu \mu$ at tree-level. 
After integrating out the LQ, one finds 
\begin{equation}
     C^{\text{BSM}} _{b_L \mu_L} = \frac{4 \pi v^2}{V_{tb} V^*_{ts} \alpha_{\rm em}}\frac{y_{3\,b\mu}^L (y_{3\,s\mu}^L)^*}{m^2_{S_3}}\, .
\end{equation}
However $S_3$ cannot address the data $R^{exp}_{D^{(*)}}>R^{SM}_{D^{(*)}}$, since it generates a negative coefficient $C_{V_L}$ (after accounting for the data on $B \to K^{(*)} \nu \bar\nu$ and $\Delta m_{B_s}$) \cite{Angelescu:2018tyl}
\begin{equation}
    C_{V_L}=-\frac{v^2}{ V_{cb}}\frac{\big( V y^{L}_3\big)^{c\tau} y^{L\ \ast}_{3\,b\tau}}{m^2_{U_3}} \ .
\end{equation}
Employing the constraint given in Eq. (\ref{Constraintbtos1}), we determine that the minimum threshold for the LQ scale is approximately $50 \, {\rm{TeV}}$.

\item{$U_1=(\mathbf{3},\mathbf{1},2/3)$}\\
$U_1$ does not couple to diquarks and its   lepto-quark interactions read:
\begin{equation}
    x^{LL}_{1\ ij}\bar{Q}^{i,a}_L\gamma^\mu U_{1,\mu}L^{j,a}_L+\ x^{RR}_{1\ ij}\bar{d}^i_R\gamma^\mu U_{1,\mu}e^j_R+{\rm{h.c.}}\ .
\end{equation}
The absence of diquark interactions, and thus of proton stability issues, render this type of LQ model appealing. 
$U_1$ generates the charged transition $b\to c \tau\bar{\nu}_{\tau}$, with the contribution 
\begin{equation}
\begin{aligned}
    C_{V_L}=&\frac{v^2}{V_{cb}}\frac{\big(Vx_{1}^{LL}\big)_{c\tau}\ x^{LL\ \ast}_{1\,b\tau}}{m^2_{U_1}}\ ,\\
 C_{S_R}=&\frac{v^2}{2V_{cb}}\frac{\big(Vx_{1}^{LL}\big)_{c\tau}\ x^{RR\ \ast}_{1\,b\tau}}{m^2_{U_1}}\ ,\\
\end{aligned}
\end{equation}
which can accomodate $R_{D^{(*)}}$.
Neglecting couplings to right-handed neutrinos, one also predicts the following Wilson's coefficients  
\begin{equation}
\begin{aligned}
 C^{\text{BSM}} _{b_L \mu_L}=-\frac{4 \pi v^2}{V_{tb}V^\ast_{ts}\alpha_{\rm em}}\frac{x^{LL}_{1\,s\mu} x^{LL\ \ast}_{1\,b\mu}}{m^2_{U_1}} \, , \\
 C^{\text{BSM}} _{b_R \mu_R}=-\frac{4 \pi v^2}{V_{tb}V^\ast_{ts}\alpha_{\rm em}}\frac{x^{RR}_{1\,s\mu} x^{RR\ \ast}_{1\,b\mu}}{m^2_{U_1}} \, .
\end{aligned}
\end{equation}
Switching on the right-handed couplings, denoted as $x_1^{RR}$, results in contributions to the Wilson coefficient $C_{b_R \mu_R}^{\mathrm{BSM}}$. As mentioned in \cref{sec:LQmotfit}, current $b \to s$ data  disfavor this Wilson coefficient. Therefore, we set the right-handed couplings to zero $x_1^{RR}=0$. This scenario is referred to as the minimal $U_1$ model, as described in \cite{Angelescu:2021lln}, where only $C_{b_L \mu_L}^{\mathrm{BSM}}$ and $C_{V_L}$ are turned on.
\\
In the natural scenario where the couplings $(x_1^{LL})^{b(c)\tau}$ and $x^{LL}_{1\,b(s)\mu}$ are all of the same order of magnitude,  the constraint in Eq. \eqref{Constraintbtos1}, from HI data for the $b\to s$ sector would require over $50~$TeV for the mass of $U_1$. However, the $b \to c$ data are not compatible with this scenario.\\ The LQ $U_1$  can accomodate both the $b \to c$ and $b \to s$ HI data under specific coupling structures, requiring a certain fine-tuning: $
    \frac{\big( V x^{LL}_1\big)_{c \tau}  x_{1\,b\tau}^{LL\ \ast}}{x^{LL}_{1\,s\mu} x^{LL\ \ast}_{1\,b\mu}}\sim 30 \ .$\\
Additionally, it is worth mentioning that the minimal $U_1$ scenario fails to account the recent findings from  Belle II experiment concerning the branching ratio $\text{BR}(B\to K \nu \nu)$, as detailed in \cite{Allwicher:2023syp}.

\item{$V_2=(\mathbf{3},\mathbf{2},5/6)$}\\
The   Lagrangian of the vector weak doublet $V_2$ is 
\begin{equation}
\begin{split}
    &x^{RL}_{2\,ij}\Bar{d}^{C\ i}_R\gamma^{\mu} V_{2,\mu}^a (i\tau^2)^{ab}L^{j,b}_L+x^{LR}_{2\,ij}\Bar{Q}^{C\ i,\,a}_L\gamma^{\mu}(i\tau^2)^{ab} V_{2,\mu}^b e^{j}_R+{\rm{h.c.}}\ .
\end{split}
\end{equation}
These interactions lead to the following contribution to the $b\to s\mu\mu$ process 
\begin{equation}
  C^{\text{BSM}} _{b_L \mu_R}=-\frac{4 \pi v^2}{V_{tb}V^\ast_{ts}\alpha_{\rm em}}\frac{x^{LR}_{2\,s\mu} x^{LR\ \ast}_{2\,b\mu}}{m^2_{V_2}} \, ,
\end{equation}
 which is compatible with the HI data. The contribution to the $b \to c$ sector is generated at tree-level:
\begin{equation}
    C_{S_R}=-\frac{2 v^2}{ V_{cb}}\frac{\big( V x^{LR}_2\big)_{c \tau} x^{RL\ \ast}_{2\,b\tau}}{m^2_{V_2}} \, .
\end{equation}

To describe both $b \to s$ and $b \to c$ HI data, in this case the tuning in the couplings should be: $
    \frac{\big( V x^{LR}_2\big)_{c \tau} x^{RL\ \ast}_{2\,b\tau}}{x^{LR}_{2\,s\mu} x^{LR\ \ast}_{2\,b\mu}}\sim 50 \ .$
    If we assume all the couplings of order one, the constraint given in Eq. (\ref{Constraintbtos2}) obtained from $b \to s$ data gives a lower bound of about $30~$TeV for the mass of $V_2$.

\item{$U_3=(\mathbf{3},\mathbf{3},2/3)$}\\
The weak triplet $U_3$  couples only to left-handed particles via the interaction terms
\begin{equation}
    x^{LL}_{3\ ij}\Bar{Q}^{i,a}_L\gamma^\mu\big(\tau^kU^k_{3, \mu}\big)^{ab}L^{j,b}_L+{\rm{h.c.}}\ .
\end{equation}
As in the $U_1$ case, diquark couplings are absent. These interactions lead to the following contribution to the $b\to s\mu\mu$ process 
\begin{equation}
  C^{\text{BSM}} _{b_L \mu_L}=-\frac{4 \pi v^2}{V_{tb}V^\ast_{ts}\alpha_{\rm em}}\frac{x^{LL}_{3\,s\mu} x^{LL\ \ast}_{3\,b\mu}}{m^2_{U_3}} \, ,
\end{equation}
 which affects the HI data in the $b \to s$ sector. 
 However, it cannot explain the experimental results for the $R_D$ and $R_{D^\ast}$, due to its negative contribution  for the $b\to c \tau\bar{\nu}_{\tau}$ process \cite{Angelescu:2018tyl}
\begin{equation}
    C_{V_L}=-\frac{v^2}{ V_{cb}}\frac{\big( V x^{LL}_3\big)^{c\tau} x^{LL\ \ast}_{3\,b\tau}}{m^2_{U_3}} \ .
\end{equation}
Utilizing the constraint presented in Eq. (\ref{Constraintbtos1}), we establish that the minimum threshold for the LQ scale, required to account for the $b \to s$ HI data, is approximately $50 \, {\rm{TeV}}$.
	\end{itemize}

We summarise our findings in table \ref{tab:Lepto}: all the leptoquark models can be compatible with the SM-like results for the $b \to s$ processes. Instead, the $b \to c$ sector can be addressed by all LQ models except for the scalar $S_3$ and the vector $U_3$. Intriguingly only the $S_1$ state can naturally fit the overall results. For $U_1$,  $V_2$ and $R_2$ a symmetry must be invoked protecting the couplings with muons. 
In the table we indicate with a $(\checkmark)$ the success in reproducing the experimental data conditioned to requiring new symmetry limits for the muonic coefficients. 

\begin{table}[h!]
\centering
\renewcommand{\arraystretch}{1.5}
\begin{tabular}{c|c|cc|c}
\toprule
LQ Model & Wilson Coeff. & $b \to c$ & $b\to s$ & $b \to c + b\to s$ \\ \hline
$S_1$          & $C^{\text{BSM}} _{b_L \mu_L}, C^{\text{BSM}} _{b_L \mu_R},C_{V_L}, C_{S_L}=-4 C_T $                      & $\checkmark$                      & $\checkmark$                      & $\checkmark$                                      \\
$R_2$          & $C^{\text{BSM}} _{b_L \mu_R}, C_{S_L}=4 C_T$                      & $\checkmark$                      & $\checkmark$                & $(\checkmark)$     \\
$S_3$          & $C^{\text{BSM}} _{b_L \mu_L},C_{V_L} $                      & $\times$                       & $\checkmark$                       & $\times$                                       \\ 
\hline
$U_1$          & $C^{\text{BSM}} _{b_L \mu_L},C_{V_L}$                      & $\checkmark$                      & $\checkmark$                      &$(\checkmark)$                                     \\
$V_2$          & $C^{\text{BSM}}_{b_L \mu_R}$, $C_{S_R}$                      & $\checkmark$                        & $\checkmark$                        & $(\checkmark)$    \\
$U_3$          & $C^{\text{BSM}} _{b_L \mu_L}$, $C_{V_L}$                     & $\times$                     &    $\checkmark$                   & $\times$    \\    
\bottomrule
\end{tabular}
\caption{Summary of leptoquark models and the relevant BSM Wilson coefficients they predict at the effective level. Checkmarks denote models successfully explaining clean observables for a particular flavor transition, while crosses signify otherwise. The final column indicates whether the model can explain both sectors simultaneously. Models marked with $(\checkmark)$ reproduce experimental data with small muonic coefficients.}
\label{tab:Lepto}
\end{table}

	\section{Conclusions}\label{conclusions}
	
 In this work we have analysed the experimental results for the neutral and charged rare $B$ meson decays, considering the latest measurements of the HI observables by the LHCb and CMS collaborations.   We started with a theoretical investigation of the $\mu/e$ ratios $R_{K}$ and  $R_{K^*}$  as well as the process $B_s \rightarrow \mu^+ \mu^-$. We studied the impact of complex Wilson coefficients and derived constraints on both their imaginary and real parts. This analysis has been then followed by a comprehensive comparison with experimental results. We find that:
	  
\begin{enumerate}
    \item The hadronic insensitive observables are currently compatible with the Standard Model prediction within $1\,\sigma$ level. 
    \item When including the hadronic sensitive observables we observe that deviations from the SM persist, with a preference of new physics in the Wilson coefficient $C^{\text{BSM}}_{b_L \mu_L}$ with a significance of $4.1\, \sigma$.

    \item Considering simultaneously all relevant Wilson coefficients and combining both hadronic sensitive and insensitive data into the fit, the deviation from the SM is observed at $4.3\,\sigma$ and decreases at the $3.3\,\sigma$, when considering the Pull value.

\end{enumerate}    
Moving on to investigate the $b\to c$ anomalies, where a violation of leptonic flavor universality is still evident in the latest measurements, we performed an analysis on the complex Wilson coefficients, establishing constraints on both their imaginary and real components. Following this analysis, the comparison with experimental results reveal that:
\begin{enumerate}
    \item Deviations from the Standard Model predictions persist, showing preference towards the emergence of a non-vanishing  real-valued NP induced coefficient $C_{V_L}$ with a significance of $4.3\,\sigma$ when solely focusing on the hadronic insensitive observables.  
    
    \item Leptoquark motivated scenarios where two Wilson coefficient are simultaneously turned on were considered, indicating a deviation at $4\,\sigma$ when employing the Pull value. This means that these models are well suited as successful extensions of the Standard Model. 

\end{enumerate}    
 Finally, we reviewed different leptoquark models aimed at explaining the deviations from the Standard Model arriving at the conclusions that: The $S_1$ leptoquark with a mass in the TeV range can naturally explain the SM-like $b \to s$ results and the anomalous deviation of the $b \to c$ processes. If, however, one is willing to accept unnaturally small couplings to the second generation leptons also the leptoquarks $U_1$, $V_2$ and $R_2$ can be employed.

Overall, even though the landscape of new physics theories has considerably shrunk some motivated extensions are still phenomenologically relevant and worth pursuing via experimental searches.  

	\subsubsection*{Acknowledgments}

The work of F.S. is
partially supported by the Carlsberg Foundation, semper ardens grant CF22-0922.\\
The work of  N.V. is supported by ICSC – Centro Nazionale di Ricerca in High Performance Computing, Big Data and Quantum Computing, funded by European Union – NextGenerationEU, reference code CN 00000013.

		\appendix

	\section{Observables}\label{appbtoc}
	In Table~\ref{tab:br_ang},\ref{tab:br_obs},\ref{tab:bcbr_obs},\ref{tab:bcbr_obs1} and \ref{tab:bcbr_tot}  we summarise the observables used in addition to the `hadronic insensitive' observables. 
	All bins are treated in the experimental analyses as independent, even if overlapping. It is clear that a correlation should 
	exists between measurements in overlapping bins, however this is not estimated by the experimental collaborations. 
	For this reason we include in our fit the measurements in all relevant bins, even if overlapping, without including any correlation 
	beyond the ones given in the experimental papers. Notice that, for instance in the case of the LHCb analysis \cite{LHCb:2015svh}, 
	the result in the bin $[1.1,6]$ GeV$^2$ has a smaller error than the measurements in the bins $[1.1,2.5],[2.5,4],[4,6]$ GeV$^2$, 
	even when the information from these three bins is combined. 
	
	\begin{table}[pt]
		\begin{center}
			\begin{tabular}{c|c}
				\toprule
				\multicolumn{2}{c}{\textbf{Angular observables}}   \\
				\midrule
				Observable  & {{$[q^{2}_{\text{min}},q^{2}_{\text{max}}]$ [GeV$^{2}$]}} \\ 
				\midrule
				\multicolumn{2}{c}{{\color{black}{LHCb $B^{+}\to K^{*+}\mu\mu$ 2020 \cite{LHCb:2020gog}, $B^0\to K^{*0}\mu\mu$ 2020 S \cite{LHCb:2020gog}}}} \\
				\midrule
				\text{$\langle F_{L} \rangle $} & [1.1, 6], [15, 19], [0.1, 0.98], [1.1, 2.5], [2.5, 4], [4, 6], [15, 17], [17, 19]  
				 \\  
				\text{$\langle S_{3} \rangle $} & [1.1, 6], [15, 19], [0.1, 0.98], [1.1, 2.5], [2.5, 4], [4, 6], [15, 17], [17, 19]  
				 \\  
				\text{$\langle S_{4} \rangle $} & [1.1, 6], [15, 19], [0.1, 0.98], [1.1, 2.5], [2.5, 4], [4, 6], [15, 17], [17, 19]  
				 \\  
				\text{$\langle S_{5} \rangle $} & \textcolor{blue}{[1.1, 6]}, [15, 19], [0.1, 0.98], [1.1, 2.5], [2.5, 4], \textcolor{blue}{[4, 6]}, [15, 17], [17, 19]  
				 \\  
				\text{$\langle S_{7} \rangle$} & [1.1, 6], [15, 19], [0.1, 0.98], [1.1, 2.5], [2.5, 4], [4, 6], [15, 17], [17, 19]  
				 \\  
				\text{$\langle S_{8} \rangle$} & [1.1, 6], [15, 19], [0.1, 0.98], \textcolor{blue}{[1.1, 2.5]}, [2.5, 4], [4, 6], [15, 17], [17, 19]  
				 \\  
				\text{$\langle S_{9} \rangle $} & [1.1, 6], [15, 19], [0.1, 0.98], [1.1, 2.5], [2.5, 4], [4, 6], [15, 17], [17, 19] \\  
				\text{$\langle  A_{FB} \rangle $} & [1.1, 6], [15, 19], [0.1, 0.98], [1.1, 2.5], [2.5, 4], [4, 6], [15, 17], [17, 19] \\  
				 \text{$\langle  P_1 \rangle $} & [1.1, 6], [15, 19], [0.1, 0.98], [1.1, 2.5], [2.5, 4], [4, 6], [15, 17], [17, 19] \\  
				\text{$\langle  P_2 \rangle $} & [1.1, 6], [15, 19], [0.1, 0.98], [1.1, 2.5], [2.5, 4], [4, 6], [15, 17], [17, 19] \\  
				\text{$\langle  P_3 \rangle $} & [1.1, 6], [15, 19], [0.1, 0.98], [1.1, 2.5], [2.5, 4], [4, 6], [15, 17], [17, 19] \\  
				\text{$\langle  P'_4 \rangle$} & [1.1, 6], [15, 19], [0.1, 0.98], [1.1, 2.5], [2.5, 4], [4, 6], [15, 17], [17, 19] \\  
				\text{$\langle  P'_5 \rangle $} & \textcolor{blue}{[1.1, 6]}, [15, 19], [0.1, 0.98], [1.1, 2.5], [2.5, 4], \textcolor{blue}{[4, 6] }, [15, 17], [17, 19] \\  
				\text{$\langle  P'_6 \rangle $} & [1.1, 6], [15, 19], [0.1, 0.98], [1.1, 2.5], [2.5, 4], [4, 6], [15, 17], [17, 19] \\  
				\text{$\langle  P'_8 \rangle$} & [1.1, 6], [15, 19], [0.1, 0.98], [1.1, 2.5], [2.5, 4], [4, 6], [15, 17], [17, 19] \\  
				\bottomrule
				\multicolumn{2}{c}{{\color{black}{CMS $B\to K^{*}\mu\mu$ 2017 \cite{CMS:2017ivg}}}} \\
				\midrule
				\text{$\langle P_{1} \rangle (B^{0}\to K^{*}\mu\mu)$} & [1, 2], [2, 4.3], [4.3, 6], [16, 19] \\  
				\text{$\langle P_{5}' \rangle (B^{0}\to K^{*}\mu\mu)$} & [1, 2], [2, 4.3], [4.3, 6], [16, 19] \\  
				\bottomrule
				\multicolumn{2}{c}{{\color{black}{ATLAS $B\to K^{*}\mu\mu$ 2017 \cite{ATLAS:2017dlm}}}} \\
				\midrule
				\text{$\langle F_{L} \rangle (B^{0}\to K^{*}\mu\mu)$} & [0.04, 2], [2, 4], [4, 6], [0.04, 4], \textcolor{blue}{[1.1, 6]}, [0.04, 6] \\  
				\text{$\langle S_{3} \rangle (B^{0}\to K^{*}\mu\mu)$} & [0.04, 2], [2, 4], [4, 6], [0.04, 4], [1.1, 6], [0.04, 6] \\  
				\text{$\langle S_{4} \rangle (B^{0}\to K^{*}\mu\mu)$} & [0.04, 2], [2, 4], \textcolor{blue}{[4, 6]}, [0.04, 4], [1.1, 6], [0.04, 6] \\  
				\text{$\langle S_{5} \rangle (B^{0}\to K^{*}\mu\mu)$} & [0.04, 2], [2, 4], \textcolor{blue}{[4, 6]}, [0.04, 4], [1.1, 6], [0.04, 6] \\  
				\text{$\langle S_{7} \rangle (B^{0}\to K^{*}\mu\mu)$} & [0.04, 2], [2, 4], [4, 6], [0.04, 4], [1.1, 6], [0.04, 6] \\  
				\text{$\langle S_{8} \rangle (B^{0}\to K^{*}\mu\mu)$} & [0.04, 2], \textcolor{blue}{[2, 4]}, [4, 6], [0.04, 4], [1.1, 6], [0.04, 6] \\  
				\text{$\langle P_{1} \rangle (B^{0}\to K^{*}\mu\mu)$} & [0.04, 2], [2, 4], [4, 6], [0.04, 4], [1.1, 6], [0.04, 6] \\  
				\text{$\langle P_{4}' \rangle (B^{0}\to K^{*}\mu\mu)$} & [0.04, 2], [2, 4], \textcolor{blue}{[4, 6]}, [0.04, 4], [1.1, 6], [0.04, 6] \\  
				\text{$\langle P_{5}' \rangle (B^{0}\to K^{*}\mu\mu)$} & [0.04, 2], [2, 4], \textcolor{blue}{[4, 6]}, [0.04, 4], [1.1, 6], [0.04, 6] \\  
				\text{$\langle P_{6}' \rangle (B^{0}\to K^{*}\mu\mu)$} & [0.04, 2], [2, 4], [4, 6], [0.04, 4], [1.1, 6], [0.04, 6] \\  
				\text{$\langle P_{8}' \rangle (B^{0}\to K^{*}\mu\mu)$} & [0.04, 2], \textcolor{blue}{[2, 4]}, [4, 6], [0.04, 4], [1.1, 6], [0.04, 6] \\  
				\bottomrule

			\end{tabular}
		\end{center}
		\caption{\em List of angular observables used in the global fit in addition to the `hadronic insensitive' observables.}
  \label{tab:br_ang}
	\end{table}%

	\normalsize
	
	\begin{table}[!htb!]
		\begin{center}
			\small\begin{tabular}{c|c}
				\toprule
				\multicolumn{2}{c}{\textbf{Branching ratios}}   \\
				\midrule
				Observable  & {{$[q^{2}_{\text{min}},q^{2}_{\text{max}}]$ [GeV$^{2}$]}} \\ \midrule
				
				\multicolumn{2}{c}{{\color{black}{LHCb $B^{\pm}\to K\mu\mu$ 2014 \cite{LHCb:2014cxe}}}} \\
				\midrule
				\multirow{2}{*}{$\frac{d}{dq^{2}}\text{BR}(B^{\pm}\to K\mu\mu)$} & [0.1, 0.98], [1.1, 2], [2, 3], [3, 4], 
				[4, 5], [5, 6], \textcolor{blue}{[15, 16]}, [16, 17],\\
				& [17, 18], [18, 19], \textcolor{red}{[19, 20]}, \textcolor{blue}{[20, 21]}, [21, 22], [1.1, 6], [15, 22]  \\   
				\bottomrule
				
				\multicolumn{2}{c}{{\color{black}{LHCb $B^{0}\to K\mu\mu$ 2014 \cite{LHCb:2014cxe}}}} \\
				\midrule
				$\frac{d}{dq^{2}}\text{BR}(B^{0}\to K\mu\mu)$ & \textcolor{blue}{[0.1, 2]}, [2, 4], \textcolor{blue}{[4, 6]}, [15, 17], [17, 22], [1.1, 6], \textcolor{blue}{[15, 22]}\\  \bottomrule
				\multicolumn{2}{c}{{\color{black}{LHCb $B^{\pm}\to K^{*}\mu\mu$ 2014 \cite{LHCb:2014cxe}}}} \\
				
				\midrule
				$\frac{d}{dq^{2}}\text{BR}(B^{\pm}\to K^{*}\mu\mu)$ & [0.1, 2], \textcolor{blue}{[2, 4]}, [4, 6], [15, 17], \textcolor{red}{[17, 19]}, [1.1, 6], 
				[15, 19] \\  \bottomrule
				
				\multicolumn{2}{c}{{\color{black}{LHCb $B^{0}\to K^{*}\mu\mu$ 2016 \cite{LHCb:2016ykl}}}} \\
				\midrule
				$\frac{d}{dq^{2}}\text{BR}(B^{0}\to K^{*}\mu\mu)$ & [0.1, 0.98], [1.1, 2.5], [2.5, 4], [4, 6], [15, 17], 
				[17, 19], [1.1, 6], [15, 19] \\  \bottomrule
				
				\multicolumn{2}{c}{{\color{black}{LHCb $B_{s}\to \phi\mu\mu$ 2021 \cite{LHCb:2021zwz}}}} \\
				\midrule
			    $\frac{d}{dq^{2}}\text{BR}(B_{s}\to \phi \mu\mu)$ & \textcolor{blue}{[0.1, 0.98]}, \textcolor{red}{[1.1, 2.5]}, \textcolor{red}{[2.5, 4]}, 
				\textcolor{blue}{[4, 6]}, [15, 17], [17, 19], \textcolor{red}{[1.1, 6]}, [15, 19] \\  \bottomrule
				
				\multicolumn{2}{c}{{\color{black}{Babar $B\to X_{s} ll$ 2013 \cite{BaBar:2013qry}}}} \\
				\midrule
				\text{$\frac{d}{dq^{2}}\text{BR} (B\to X_{s} ll)$} &\textcolor{blue}{[0.1, 2]}, [2.0, 4.3], [4.3, 6.8],  [1, 6], [14,2, 25] \\  
				\text{$\frac{d}{dq^{2}}\text{BR}(B\to X_{s}\mu\mu)$} &[0.1, 2], [2.0, 4.3], [4.3, 6.8],  [1, 6], [14,2, 25]\\ 
				\text{$\frac{d}{dq^{2}}\text{BR}(B\to X_{s} ee)$}&\textcolor{blue}{[0.1, 2]}, [2.0, 4.3], [4.3, 6.8],  [1, 6], [14,2, 25]\\ 
				\bottomrule
				\multicolumn{2}{c}{{\color{black}{Belle $B \to X_{s}ll$ 2005 \cite{Belle:2005fli}}}} \\
				\midrule
				$\frac{d}{dq^{2}}\text{BR}(B\to X_{s} ll)$ & [0.04, 1], [1, 6], [14.4, 25]\\  \bottomrule
				
			\end{tabular}
		\end{center}
		\caption{\em List of differential branching ratios used in the global fit in addition to the `hadronic insensitive' observables. The bins highlighted in red refer to measurements which deviate more than $3.5$ $\sigma$ from the theoretical prediction.
			\label{tab:br_obs}}
	\end{table}%
	\normalsize

		\begin{table}[!htb!]
		\begin{center}
			\small\begin{tabular}{c|c}
				\toprule
				\multicolumn{2}{c}{\textbf{Binned Branching ratios}}   \\
				\midrule
				Observable  & {{$[q^{2}_{\text{min}},q^{2}_{\text{max}}]$ [GeV$^{2}$]}} \\ \midrule
				
				\multicolumn{2}{c}{{\color{black}{Belle $B^{+}\to D\mu\nu_\mu$ 2015\cite{Belle:2015pkj}}}} \\
				\midrule
				\multirow{2}{*}{$\text{BR}(B^{+}\to D\mu\nu_\mu)$} & [0.0, 1.03], [1.03, 2.21], [2.21, 3.39], [3.39, 4.57], 
				[4.57, 5.75], \\& [5.75, 6.93], [6.93, 8.11], [8.11, 9.3],
				 [9.3, 10.48], [10.48, 11.66]\\   
				\bottomrule
				
				\multicolumn{2}{c}{{\color{black}{Belle $B^{0}\to D\mu\nu_\mu$ 2015 \cite{Belle:2015pkj}}}} \\
				\midrule
				$\text{BR}(B^{0}\to D\mu\nu_\mu)$ &[0.0, 1.03], [1.03, 2.21], [2.21, 3.39], [3.39, 4.57], 
				[4.57, 5.75], \\&[5.75, 6.93],[6.93, 8.11], [8.11, 9.3],
				 [9.3, 10.48], [10.48, 11.66]\\  \bottomrule
				\multicolumn{2}{c}{{\color{black}{Belle $B^{+}\to D e\nu_e$ 2015 \cite{Belle:2015pkj}}}} \\
				
				\midrule
				$\text{BR}(B^{+}\to D e\nu_e)$ & [0.0, 1.03], [1.03, 2.21], [2.21, 3.39], [3.39, 4.57], 
				[4.57, 5.75], \\& [5.75, 6.93],[6.93, 8.11], [8.11, 9.3], [9.3, 10.48], [10.48, 11.66] \\  \bottomrule
				\multicolumn{2}{c}{{\color{black}{Belle $B^{0}\to D e\nu_e$ 2015 \cite{Belle:2015pkj}}}} \\
				\midrule
				$\text{BR}(B^{0}\to D e\nu_e)$ & [0.0, 1.03], [1.03, 2.21], [2.21, 3.39], [3.39, 4.57], 
				[4.57, 5.75],\\& [5.75, 6.93],[6.93, 8.11], [8.11, 9.3], [9.3, 10.48], [10.48, 11.66] \\  \bottomrule
				\multicolumn{2}{c}{{\color{black}{Babar $B^+\to D \ell \nu_\ell$ 2009 \cite{BaBar:2009zxk}}}} \\
				\midrule
			    $\text{BR}(B^+\to D \ell \nu_\ell)$ & [0.0, 0.97], [0.97, 2.15], [2.15, 3.34], [3.34, 4.52], 
				[4.52, 5.71],\\& [5.71, 6.89], [6.89, 8.07], [8.07, 9.26], [9.26, 10.44], [10.44, 11.63] \\ \bottomrule
				
				\multicolumn{2}{c}{{\color{black}{Belle $B \to D^* \mu \nu_\mu$ 2010 \cite{Belle:2010qug}}}} \\
				\midrule
				$\text{BR}_L(B\to D^* \mu \nu_\mu)$ & [0.08, 1.14], [1.14, 2.2], [2.2, 3.26], [3.26, 4.32], 
				[4.32, 5.38],\\& [5.38, 6.44], [6.44, 7.5], [7.5, 8.57], [8.57, 9.63], [9.63, 10.69]\\  
    $\text{BR}_T(B\to D^* \mu \nu_\mu)$ & [0.08, 1.14], [1.14, 2.2], [2.2, 3.26], [3.26, 4.32], 
				[4.32, 5.38],\\& [5.38, 6.44], [6.44, 7.5], [7.5, 8.57], [8.57, 9.63], [9.63, 10.69]\\  
    \bottomrule
    \multicolumn{2}{c}{{\color{black}{Belle $B \to D^* e \nu_e$ 2010 \cite{Belle:2010qug}}}} \\
				\midrule
				$\text{BR}_L(B\to D^* e\nu_e)$ & [0.08, 1.14], [1.14, 2.2], [2.2, 3.26], [3.26, 4.32], 
				[4.32, 5.38], \\&[5.38, 6.44], [6.44, 7.5], [7.5, 8.57], [8.57, 9.63], [9.63, 10.69]\\ 
    $\text{BR}_T(B\to D^* e \nu_e)$ & [0.08, 1.14], [1.14, 2.2], [2.2, 3.26], [3.26, 4.32], 
				[4.32, 5.38],\\& [5.38, 6.44], [6.44, 7.5], [7.5, 8.57], [8.57, 9.63], [9.63, 10.69]\\  \bottomrule
\multicolumn{2}{c}{{\color{black}{Belle $B^0\to D^* \ell \nu_\ell$ 2017 \cite{Belle:2017rcc}}}} \\
				\midrule
				$\text{BR}(B^0\to D^* \ell \nu_\ell)$ & [0.0, 1.14], [1.14, 2.2], [2.2, 3.26], [3.26, 4.32], 
				[4.32, 5.38], \\&[5.38, 6.44], [6.44, 7.5], [7.5, 8.57], [8.57, 9.63], [9.63, 10.69]  \\  
				\bottomrule				
			\end{tabular}
		\end{center}
		\caption{\em List of differential branching ratios used in the global fit for the $b\to c$ sector. The bins highlighted in red refer to measurements which deviate more than $3.5$ $\sigma$ from the theoretical prediction.
			\label{tab:bcbr_obs}}
	\end{table}%
	\normalsize
\begin{table}[!htb!]
		\begin{center}
			\small\begin{tabular}{c|c}
				\toprule
				\multicolumn{2}{c}{\textbf{Binned Branching ratios}}   \\
				\midrule
				Observable  & {{$[\cos{\theta}_v^{\text{min}},\cos{\theta}_v^{\text{max}}]$}} \cite{Belle:2017rcc}\\ \midrule
    $\text{BR}(B\to D^* \ell \nu_\ell)$ & [-1.0, -0.8], [-0.8,-0.6],[-0.6,-0.4], [-0.4,-0.2],[-0.2,0.0], \\&[0.2,0.4]
				,[0.4,0.6],[0.6,0.8], [0.8,1.0]\\  \midrule
\multicolumn{2}{c}{{\color{black}{ {{$[\cos{\theta}_l^{\text{min}},\cos{\theta}_l^{\text{max}}]$}}}\cite{Belle:2017rcc}}} \\
				\midrule
    $\text{BR}(B\to D^* \ell \nu_\ell)$ & [-1.0, -0.8], [-0.8,-0.6],[-0.6,-0.4], [-0.4,-0.2],[-0.2,0.0], \\&[0.2,0.4],[0.4,0.6],[0.6,0.8],
				[0.8,1.0]\\  \midrule
  				 \multicolumn{2}{c}{{\color{black}{ {{$[\phi^{\text{min}},\phi^{\text{max}}]$}}}\cite{Belle:2017rcc}}}\\
				\midrule
    $\text{BR}(B\to D^* \ell \nu_\ell)$ &$[0.0,\pi/5] ,[\pi/5,2\pi/5]
				,[2\pi/5,3\pi/5],[3\pi/5,4\pi/5]
				,[4\pi/5,\pi],$\\&
				$[\pi,6\pi/5]
				,[6\pi/5,7\pi/5]
				,[7\pi/5,8\pi/5],[8\pi/5,9\pi/5]
				,[9\pi/5,2\pi]$
				\\  \bottomrule
				
			\end{tabular}
		\end{center}
		\caption{\em List of differential branching ratios used in the global fit for the $b\to c$ sector. The bins highlighted in red refer to measurements which deviate more than $3.5$ $\sigma$ from the theoretical prediction.
			\label{tab:bcbr_obs1}}
	\end{table}%

 \normalsize
\begin{table}[!htb!]
		\begin{center}
			\small\begin{tabular}{c|c}
				\toprule
				\multicolumn{2}{c}{\textbf{Total Branching ratios} \cite{Adamczyk:2019wyt,HFLAV:2016hnz,BaBar:2008zui,BaBar:2007cke,BaBar:2007ddh,BaBar:2007nwi}}   \\
				\midrule
				Observable  & {{$\text{BR}(B^+\to D e \nu_e)$\ , $\text{BR}(B^+\to D^* e \nu_e)$ \ , $\text{BR}(B^+\to D \mu \nu_\mu)$}} \\  & {{$\text{BR}(B^+\to D^* \mu \nu_\mu)$ \ , $\text{BR}(B^+\to D^* \ell \nu_\ell)$\ , $\text{BR}(B^0\to D^* \ell \nu_\ell)$}}
				\\  \bottomrule
			\end{tabular}
		\end{center}
		\caption{\em List of branching ratios used in the global fit for the $b\to c$ sector. The bins highlighted in red refer to measurements which deviate more than $3.5$ $\sigma$ from the theoretical prediction.
			\label{tab:bcbr_tot}}
	\end{table}
 	\FloatBarrier
	 \printbibliography	

@article{LHCb:2022qnv,
    author = "Aaij, R. and others",
    collaboration = "LHCb",
    title = "{Test of lepton universality in $b \rightarrow s \ell^+ \ell^-$ decays}",
    eprint = "2212.09152",
    archivePrefix = "arXiv",
    primaryClass = "hep-ex",
    reportNumber = "LHCb-PAPER-2022-046, CERN-EP-2022-277",
    doi = "10.1103/PhysRevLett.131.051803",
    journal = "Phys. Rev. Lett.",
    volume = "131",
    number = "5",
    pages = "051803",
    year = "2023"
}

@article{LHCb:2022zom,
    author = "Aaij, R. and others",
    collaboration = "LHCb",
    title = "{Measurement of lepton universality parameters in $B^+\to K^+\ell^+\ell^-$ and $B^0\to K^{*0}\ell^+\ell^-$ decays}",
    eprint = "2212.09153",
    archivePrefix = "arXiv",
    primaryClass = "hep-ex",
    reportNumber = "LHCb-PAPER-2022-045, CERN-EP-2022-278",
    doi = "10.1103/PhysRevD.108.032002",
    journal = "Phys. Rev. D",
    volume = "108",
    number = "3",
    pages = "032002",
    year = "2023"
}

@article{CMS:2022mgd,
    author = "Tumasyan, Armen and others",
    collaboration = "CMS",
    title = "{Measurement of the B$^0_\mathrm{S}$$\to$$\mu^+\mu^-$ decay properties and search for the B$^0$$\to$$\mu^+\mu^-$ decay in proton-proton collisions at $\sqrt{s}$ = 13 TeV}",
    eprint = "2212.10311",
    archivePrefix = "arXiv",
    primaryClass = "hep-ex",
    reportNumber = "CMS-BPH-21-006, CERN-EP-2022-270",
    doi = "10.1016/j.physletb.2023.137955",
    journal = "Phys. Lett. B",
    volume = "842",
    pages = "137955",
    year = "2023"
}

@article{HeavyFlavorAveragingGroup:2022wzx,
    author = "Amhis, Yasmine Sara and others",
    collaboration = "HFLAV",
    title = "{Averages of b-hadron, c-hadron, and \ensuremath{\tau}-lepton properties as of 2021}",
    eprint = "2206.07501",
    archivePrefix = "arXiv",
    primaryClass = "hep-ex",
    doi = "10.1103/PhysRevD.107.052008",
    journal = "Phys. Rev. D",
    volume = "107",
    number = "5",
    pages = "052008",
    year = "2023"
}

@article{LHCb:2014cxe,
    author = "Aaij, R. and others",
    collaboration = "LHCb",
    title = "{Differential branching fractions and isospin asymmetries of $B \to K^{(*)} \mu^+ \mu^-$ decays}",
    eprint = "1403.8044",
    archivePrefix = "arXiv",
    primaryClass = "hep-ex",
    reportNumber = "LHCB-PAPER-2014-006, CERN-PH-EP-2014-055",
    doi = "10.1007/JHEP06(2014)133",
    journal = "JHEP",
    volume = "06",
    pages = "133",
    year = "2014"
}

@article{BELLE:2019xld,
    author = "Choudhury, S. and others",
    collaboration = "BELLE",
    title = "{Test of lepton flavor universality and search for lepton flavor violation in $B \rightarrow K\ell \ell$ decays}",
    eprint = "1908.01848",
    archivePrefix = "arXiv",
    primaryClass = "hep-ex",
    reportNumber = "BELLE-CONF-1904, Belle Preprint 2020-11, KEK Preprint 2020-12",
    doi = "10.1007/JHEP03(2021)105",
    journal = "JHEP",
    volume = "03",
    pages = "105",
    year = "2021"
}

@article{LHCb:2017avl,
    author = "Aaij, R. and others",
    collaboration = "LHCb",
    title = "{Test of lepton universality with $B^{0} \rightarrow K^{*0}\ell^{+}\ell^{-}$ decays}",
    eprint = "1705.05802",
    archivePrefix = "arXiv",
    primaryClass = "hep-ex",
    reportNumber = "LHCB-PAPER-2017-013, CERN-EP-2017-100",
    doi = "10.1007/JHEP08(2017)055",
    journal = "JHEP",
    volume = "08",
    pages = "055",
    year = "2017"
}

@article{LHCb:2021trn,
    author = "Aaij, Roel and others",
    collaboration = "LHCb",
    title = "{Test of lepton universality in beauty-quark decays}",
    eprint = "2103.11769",
    archivePrefix = "arXiv",
    primaryClass = "hep-ex",
    reportNumber = "LHCb-PAPER-2021-004, CERN-EP-2021-042",
    doi = "10.1038/s41567-023-02095-3",
    journal = "Nature Phys.",
    volume = "18",
    number = "3",
    pages = "277--282",
    year = "2022",
    note = "[Addendum: Nature Phys. 19, (2023)]"
}

@article{LHCb:2021awg,
    author = "Aaij, Roel and others",
    collaboration = "LHCb",
    title = "{Measurement of the $B^0_s\to\mu^+\mu^-$ decay properties and search for the $B^0\to\mu^+\mu^-$ and $B^0_s\to\mu^+\mu^-\gamma$ decays}",
    eprint = "2108.09283",
    archivePrefix = "arXiv",
    primaryClass = "hep-ex",
    reportNumber = "CERN-EP-2021-133, LHCb-PAPER-2021-008",
    doi = "10.1103/PhysRevD.105.012010",
    journal = "Phys. Rev. D",
    volume = "105",
    number = "1",
    pages = "012010",
    year = "2022"
}

@article{LHCb:2021zwz,
    author = "Aaij, Roel and others",
    collaboration = "LHCb",
    title = "{Branching Fraction Measurements of the Rare $B^0_s\rightarrow\phi\mu^+\mu^-$ and $B^0_s\rightarrow f_2^\prime(1525)\mu^+\mu^-$- Decays}",
    eprint = "2105.14007",
    archivePrefix = "arXiv",
    primaryClass = "hep-ex",
    reportNumber = "LHCb-PAPER-2021-014, CERN-EP-2021-092",
    doi = "10.1103/PhysRevLett.127.151801",
    journal = "Phys. Rev. Lett.",
    volume = "127",
    number = "15",
    pages = "151801",
    year = "2021"
}

@article{LHCb:2020gog,
    author = "Aaij, Roel and others",
    collaboration = "LHCb",
    title = "{Angular Analysis of the  $B^{+}\rightarrow K^{\ast+}\mu^{+}\mu^{-}$ Decay}",
    eprint = "2012.13241",
    archivePrefix = "arXiv",
    primaryClass = "hep-ex",
    reportNumber = "LHCb-PAPER-2020-041, CERN-EP-2020-239",
    doi = "10.1103/PhysRevLett.126.161802",
    journal = "Phys. Rev. Lett.",
    volume = "126",
    number = "16",
    pages = "161802",
    year = "2021"
}

@article{Belle:2019oag,
    author = "Abdesselam, A. and others",
    collaboration = "Belle",
    title = "{Test of Lepton-Flavor Universality in ${B\to K^\ast\ell^+\ell^-}$ Decays at Belle}",
    eprint = "1904.02440",
    archivePrefix = "arXiv",
    primaryClass = "hep-ex",
    reportNumber = "BELLE-CONF-1901, Belle Preprint 2020-14, KEK Preprint 2020-16",
    doi = "10.1103/PhysRevLett.126.161801",
    journal = "Phys. Rev. Lett.",
    volume = "126",
    number = "16",
    pages = "161801",
    year = "2021"
}

@article{ATLAS:2018cur,
    author = "Aaboud, Morad and others",
    collaboration = "ATLAS",
    title = "{Study of the rare decays of $B^0_s$ and $B^0$ mesons into muon pairs using data collected during 2015 and 2016 with the ATLAS detector}",
    eprint = "1812.03017",
    archivePrefix = "arXiv",
    primaryClass = "hep-ex",
    reportNumber = "CERN-EP-2018-291",
    doi = "10.1007/JHEP04(2019)098",
    journal = "JHEP",
    volume = "04",
    pages = "098",
    year = "2019"
}

@article{CMS:2019qnb,
    collaboration = "CMS",
    title = "{Measurement of properties of Bs0 to mu+mu- decays and search for
B0 to mu+mu- with the CMS experiment}",
    reportNumber = "CMS-PAS-BPH-16-004, CMS-PAS-BPH-16-004",
    year = "2019"
}

@article{LHCb:2016ykl,
    author = "Aaij, Roel and others",
    collaboration = "LHCb",
    title = "{Measurements of the S-wave fraction in $B^{0}\rightarrow K^{+}\pi^{-}\mu^{+}\mu^{-}$ decays and the $B^{0}\rightarrow K^{\ast}(892)^{0}\mu^{+}\mu^{-}$ differential branching fraction}",
    eprint = "1606.04731",
    archivePrefix = "arXiv",
    primaryClass = "hep-ex",
    reportNumber = "CERN-EP-2016-141, LHCB-PAPER-2016-012",
    doi = "10.1007/JHEP11(2016)047",
    journal = "JHEP",
    volume = "11",
    pages = "047",
    year = "2016",
    note = "[Erratum: JHEP 04, 142 (2017)]"
}

@article{BaBar:2013qry,
    author = "Lees, J. P. and others",
    collaboration = "BaBar",
    title = "{Measurement of the $B \to X_s l^+l^-$ branching fraction and search for direct CP violation from a sum of exclusive final states}",
    eprint = "1312.5364",
    archivePrefix = "arXiv",
    primaryClass = "hep-ex",
    reportNumber = "BABAR-PUB-13-01, SLAC-PUB-15866",
    doi = "10.1103/PhysRevLett.112.211802",
    journal = "Phys. Rev. Lett.",
    volume = "112",
    pages = "211802",
    year = "2014"
}

@article{Belle:2005fli,
    author = "Iwasaki, M. and others",
    collaboration = "Belle",
    title = "{Improved measurement of the electroweak penguin process $B \to X_s l^+ l^-$}",
    eprint = "hep-ex/0503044",
    archivePrefix = "arXiv",
    reportNumber = "KEK-PREPRINT-2005-5, BELLE-PREPRINT-2005-14",
    doi = "10.1103/PhysRevD.72.092005",
    journal = "Phys. Rev. D",
    volume = "72",
    pages = "092005",
    year = "2005"
}

@article{CMS:2017ivg,
    collaboration = "CMS",
    title = "{Measurement of the $P_1$ and $P_5'$ angular parameters of the decay $\mathrm{B}^0 \to \mathrm{K}^{*0} \mu^+ \mu^-$ in proton-proton collisions at $\sqrt{s}=8~\mathrm{TeV}$}",
    reportNumber = "CMS-PAS-BPH-15-008",
    year = "2017"
}

@article{ATLAS:2017dlm,
    editor = "Auge, Etienne and Dumarchez, Jacques and Tran Thanh Van, Jean",
    collaboration = "ATLAS",
    title = "{Angular analysis of $B^0_d \to K^{*}\mu^+\mu^-$ decays in $pp$ collisions at $\sqrt{s}= 8$ TeV with the ATLAS detector}",
    reportNumber = "ATLAS-CONF-2017-023",
    month = "4",
    year = "2017"
}

@article{DAlise:2022ypp,
    author = "D'Alise, Alessandra and others",
    title = "{Standard model anomalies: lepton flavour non-universality, g \ensuremath{-} 2 and W-mass}",
    eprint = "2204.03686",
    archivePrefix = "arXiv",
    primaryClass = "hep-ph",
    reportNumber = "CERN-TH-2022-063",
    doi = "10.1007/JHEP08(2022)125",
    journal = "JHEP",
    volume = "08",
    pages = "125",
    year = "2022"
}

@article{Straub:2018kue,
    author = "Straub, David M.",
    title = "{flavio: a Python package for flavour and precision phenomenology in the Standard Model and beyond}",
    eprint = "1810.08132",
    archivePrefix = "arXiv",
    primaryClass = "hep-ph",
    month = "10",
    year = "2018"
}

@article{Aebischer:2017gaw,
    author = "Aebischer, Jason and Fael, Matteo and Greub, Christoph and Virto, Javier",
    title = "{B physics Beyond the Standard Model at One Loop: Complete Renormalization Group Evolution below the Electroweak Scale}",
    eprint = "1704.06639",
    archivePrefix = "arXiv",
    primaryClass = "hep-ph",
    doi = "10.1007/JHEP09(2017)158",
    journal = "JHEP",
    volume = "09",
    pages = "158",
    year = "2017"
}

@article{Aebischer:2018bkb,
    author = "Aebischer, Jason and Kumar, Jacky and Straub, David M.",
    title = "{Wilson: a Python package for the running and matching of Wilson coefficients above and below the electroweak scale}",
    eprint = "1804.05033",
    archivePrefix = "arXiv",
    primaryClass = "hep-ph",
    doi = "10.1140/epjc/s10052-018-6492-7",
    journal = "Eur. Phys. J. C",
    volume = "78",
    number = "12",
    pages = "1026",
    year = "2018"
}

@article{Alonso:2014csa,
    author = "Alonso, Rodrigo and Grinstein, Benjamin and Martin Camalich, Jorge",
    title = "{$SU(2)\times U(1)$ gauge invariance and the shape of new physics in rare $B$ decays}",
    eprint = "1407.7044",
    archivePrefix = "arXiv",
    primaryClass = "hep-ph",
    doi = "10.1103/PhysRevLett.113.241802",
    journal = "Phys. Rev. Lett.",
    volume = "113",
    pages = "241802",
    year = "2014"
}

@article{DAmico:2017mtc,
    author = "D'Amico, Guido and Nardecchia, Marco and Panci, Paolo and Sannino, Francesco and Strumia, Alessandro and Torre, Riccardo and Urbano, Alfredo",
    title = "{Flavour anomalies after the $R_{K^*}$ measurement}",
    eprint = "1704.05438",
    archivePrefix = "arXiv",
    primaryClass = "hep-ph",
    reportNumber = "CP3-ORIGINS-2017-014, CERN-TH-2017-086, IFUP-TH/2017, CP3-Origins-2017-014",
    doi = "10.1007/JHEP09(2017)010",
    journal = "JHEP",
    volume = "09",
    pages = "010",
    year = "2017"
}

@article{Ghosh:2014awa,
    author = "Ghosh, Diptimoy and Nardecchia, Marco and Renner, S. A.",
    title = "{Hint of Lepton Flavour Non-Universality in $B$ Meson Decays}",
    eprint = "1408.4097",
    archivePrefix = "arXiv",
    primaryClass = "hep-ph",
    doi = "10.1007/JHEP12(2014)131",
    journal = "JHEP",
    volume = "12",
    pages = "131",
    year = "2014"
}

@article{Altmannshofer:2014rta,
    author = "Altmannshofer, Wolfgang and Straub, David M.",
    title = "{New physics in $b\rightarrow s$ transitions after LHC run 1}",
    eprint = "1411.3161",
    archivePrefix = "arXiv",
    primaryClass = "hep-ph",
    doi = "10.1140/epjc/s10052-015-3602-7",
    journal = "Eur. Phys. J. C",
    volume = "75",
    number = "8",
    pages = "382",
    year = "2015"
}

@article{Descotes-Genon:2015uva,
    author = "Descotes-Genon, S\'ebastien and Hofer, Lars and Matias, Joaquim and Virto, Javier",
    title = "{Global analysis of $b\to s\ell\ell$ anomalies}",
    eprint = "1510.04239",
    archivePrefix = "arXiv",
    primaryClass = "hep-ph",
    reportNumber = "LPT-ORSAY-15-68, QFET-2015-29, SI-HEP-2015-19",
    doi = "10.1007/JHEP06(2016)092",
    journal = "JHEP",
    volume = "06",
    pages = "092",
    year = "2016"
}

@article{Altmannshofer:2013foa,
    author = "Altmannshofer, Wolfgang and Straub, David M.",
    title = "{New Physics in $B \to K^*\mu\mu$?}",
    eprint = "1308.1501",
    archivePrefix = "arXiv",
    primaryClass = "hep-ph",
    reportNumber = "FERMILAB-PUB-13-310-T, MITP-13-047",
    doi = "10.1140/epjc/s10052-013-2646-9",
    journal = "Eur. Phys. J. C",
    volume = "73",
    pages = "2646",
    year = "2013"
}

@article{Hurth:2013ssa,
    author = "Hurth, Tobias and Mahmoudi, Farvah",
    title = "{On the LHCb anomaly in B $\to K^*\ell^+\ell^-$}",
    eprint = "1312.5267",
    archivePrefix = "arXiv",
    primaryClass = "hep-ph",
    reportNumber = "CERN-PH-TH-2013-256, MITP-13-064",
    doi = "10.1007/JHEP04(2014)097",
    journal = "JHEP",
    volume = "04",
    pages = "097",
    year = "2014"
}

@article{Hiller:2014yaa,
    author = "Hiller, Gudrun and Schmaltz, Martin",
    title = "{$R_K$ and future $b \to s \ell \ell$ physics beyond the standard model opportunities}",
    eprint = "1408.1627",
    archivePrefix = "arXiv",
    primaryClass = "hep-ph",
    reportNumber = "DO-TH-14-17",
    doi = "10.1103/PhysRevD.90.054014",
    journal = "Phys. Rev. D",
    volume = "90",
    pages = "054014",
    year = "2014"
}

@article{Hurth:2014vma,
    author = "Hurth, T. and Mahmoudi, F. and Neshatpour, S.",
    title = "{Global fits to $b \to s\ell\ell$ data and signs for lepton non-universality}",
    eprint = "1410.4545",
    archivePrefix = "arXiv",
    primaryClass = "hep-ph",
    reportNumber = "CERN-PH-TH-2014-198, MITP-14-078",
    doi = "10.1007/JHEP12(2014)053",
    journal = "JHEP",
    volume = "12",
    pages = "053",
    year = "2014"
}

@article{Ciuchini:2017mik,
    author = "Ciuchini, Marco and Coutinho, Antonio M. and Fedele, Marco and Franco, Enrico and Paul, Ayan and Silvestrini, Luca and Valli, Mauro",
    title = "{On Flavourful Easter eggs for New Physics hunger and Lepton Flavour Universality violation}",
    eprint = "1704.05447",
    archivePrefix = "arXiv",
    primaryClass = "hep-ph",
    doi = "10.1140/epjc/s10052-017-5270-2",
    journal = "Eur. Phys. J. C",
    volume = "77",
    number = "10",
    pages = "688",
    year = "2017"
}

@article{Capdevila:2017bsm,
    author = "Capdevila, Bernat and Crivellin, Andreas and Descotes-Genon, S\'ebastien and Matias, Joaquim and Virto, Javier",
    title = "{Patterns of New Physics in $b\to s\ell^+\ell^-$ transitions in the light of recent data}",
    eprint = "1704.05340",
    archivePrefix = "arXiv",
    primaryClass = "hep-ph",
    reportNumber = "PSI-PR-17-05, LPT-ORSAY-17-19",
    doi = "10.1007/JHEP01(2018)093",
    journal = "JHEP",
    volume = "01",
    pages = "093",
    year = "2018"
}

@article{Altmannshofer:2017fio,
    author = "Altmannshofer, Wolfgang and Niehoff, Christoph and Stangl, Peter and Straub, David M.",
    title = "{Status of the $B\rightarrow K^*\mu ^+\mu ^-$ anomaly after Moriond 2017}",
    eprint = "1703.09189",
    archivePrefix = "arXiv",
    primaryClass = "hep-ph",
    doi = "10.1140/epjc/s10052-017-4952-0",
    journal = "Eur. Phys. J. C",
    volume = "77",
    number = "6",
    pages = "377",
    year = "2017"
}

@article{Alguero:2019ptt,
    author = "Alguer\'o, Marcel and Capdevila, Bernat and Crivellin, Andreas and Descotes-Genon, S\'ebastien and Masjuan, Pere and Matias, Joaquim and Novoa Brunet, Mart\'\i{}n and Virto, Javier",
    title = "{Emerging patterns of New Physics with and without Lepton Flavour Universal contributions}",
    eprint = "1903.09578",
    archivePrefix = "arXiv",
    primaryClass = "hep-ph",
    reportNumber = "PSI-PR-19-04, ZH-TH-14/19, TUM-HEP-1192/19, MIT-CTP/5107,
  LPT-Orsay-19-11",
    doi = "10.1140/epjc/s10052-019-7216-3",
    journal = "Eur. Phys. J. C",
    volume = "79",
    number = "8",
    pages = "714",
    year = "2019",
    note = "[Addendum: Eur.Phys.J.C 80, 511 (2020)]"
}

@article{Aebischer:2019mlg,
    author = "Aebischer, Jason and Altmannshofer, Wolfgang and Guadagnoli, Diego and Reboud, M\'eril and Stangl, Peter and Straub, David M.",
    title = "{$B$-decay discrepancies after Moriond 2019}",
    eprint = "1903.10434",
    archivePrefix = "arXiv",
    primaryClass = "hep-ph",
    doi = "10.1140/epjc/s10052-020-7817-x",
    journal = "Eur. Phys. J. C",
    volume = "80",
    number = "3",
    pages = "252",
    year = "2020"
}

@article{Ciuchini:2020gvn,
    author = "Ciuchini, Marco and Fedele, Marco and Franco, Enrico and Paul, Ayan and Silvestrini, Luca and Valli, Mauro",
    title = "{Lessons from the $B^{0,+}\to K^{*0,+}\mu^+\mu^-$ angular analyses}",
    eprint = "2011.01212",
    archivePrefix = "arXiv",
    primaryClass = "hep-ph",
    reportNumber = "DESY 20-190, DESY-20-190, HU-EP-20/30, HU 20/30, TTP20-037, P3H-20-064, UCI-TR 2020-18",
    doi = "10.1103/PhysRevD.103.015030",
    journal = "Phys. Rev. D",
    volume = "103",
    number = "1",
    pages = "015030",
    year = "2021"
}

@article{Hiller:2003js,
    author = "Hiller, Gudrun and Kruger, Frank",
    title = "{More model-independent analysis of $b \to s$ processes}",
    eprint = "hep-ph/0310219",
    archivePrefix = "arXiv",
    reportNumber = "LMU-18-03, TUM-HEP-519-03",
    doi = "10.1103/PhysRevD.69.074020",
    journal = "Phys. Rev. D",
    volume = "69",
    pages = "074020",
    year = "2004"
}

@article{Bobeth:2008ij,
    author = "Bobeth, Christoph and Hiller, Gudrun and Piranishvili, Giorgi",
    title = "{CP Asymmetries in bar $B \to \bar{K}^* (\to \bar{K} \pi) \bar{\ell} \ell$ and Untagged $\bar{B}_s$, $B_s \to \phi (\to K^{+} K^-) \bar{\ell} \ell$ Decays at NLO}",
    eprint = "0805.2525",
    archivePrefix = "arXiv",
    primaryClass = "hep-ph",
    reportNumber = "DO-TH-08-03",
    doi = "10.1088/1126-6708/2008/07/106",
    journal = "JHEP",
    volume = "07",
    pages = "106",
    year = "2008"
}

@article{Hambrock:2013zya,
    author = "Hambrock, Christian and Hiller, Gudrun and Schacht, Stefan and Zwicky, Roman",
    title = "{$B \to K^\star$ form factors from flavor data to QCD and back}",
    eprint = "1308.4379",
    archivePrefix = "arXiv",
    primaryClass = "hep-ph",
    reportNumber = "DO-TH-13-13, QFET-2013-04, EDINBURGH-13-17, CP\textasciicircum{}3-ORIGINS-2013-025, DIAS-2013-25",
    doi = "10.1103/PhysRevD.89.074014",
    journal = "Phys. Rev. D",
    volume = "89",
    number = "7",
    pages = "074014",
    year = "2014"
}

@article{Hiller:2014ula,
    author = "Hiller, Gudrun and Schmaltz, Martin",
    title = "{Diagnosing lepton-nonuniversality in $b \to s \ell \ell$}",
    eprint = "1411.4773",
    archivePrefix = "arXiv",
    primaryClass = "hep-ph",
    reportNumber = "DO-TH-14-25, QFET-2014-21",
    doi = "10.1007/JHEP02(2015)055",
    journal = "JHEP",
    volume = "02",
    pages = "055",
    year = "2015"
}

@article{Allanach:2022iod,
    author = "Allanach, Ben and Davighi, Joe",
    title = "{The Rumble in the Meson: a leptoquark versus a Z' to fit b \textrightarrow{} s\ensuremath{\mu}$^{+}$\ensuremath{\mu}$^{-}$ anomalies including 2022 LHCb $ {R}_{K^{\left(\ast \right)}} $ measurements}",
    eprint = "2211.11766",
    archivePrefix = "arXiv",
    primaryClass = "hep-ph",
    reportNumber = "CERN-TH-2022-181",
    doi = "10.1007/JHEP04(2023)033",
    journal = "JHEP",
    volume = "04",
    pages = "033",
    year = "2023"
}

@article{Blanke:2018yud,
    author = "Blanke, Monika and Crivellin, Andreas and de Boer, Stefan and Kitahara, Teppei and Moscati, Marta and Nierste, Ulrich and Ni\v{s}and\v{z}i\'c, Ivan",
    title = "{Impact of polarization observables and $ B_c\to \tau \nu$ on new physics explanations of the $b\to c \tau \nu$ anomaly}",
    eprint = "1811.09603",
    archivePrefix = "arXiv",
    primaryClass = "hep-ph",
    reportNumber = "PSI-PR-18-16; TTP-18-42, PSI-PR--18--16, TTP--18--42",
    doi = "10.1103/PhysRevD.99.075006",
    journal = "Phys. Rev. D",
    volume = "99",
    number = "7",
    pages = "075006",
    year = "2019"
}

@article{Angelescu:2021lln,
    author = "Angelescu, Andrei and Be\v{c}irevi\'c, Damir and Faroughy, Darius A. and Jaffredo, Florentin and Sumensari, Olcyr",
    title = "{Single leptoquark solutions to the B-physics anomalies}",
    eprint = "2103.12504",
    archivePrefix = "arXiv",
    primaryClass = "hep-ph",
    reportNumber = "ZU-TH 12/21",
    doi = "10.1103/PhysRevD.104.055017",
    journal = "Phys. Rev. D",
    volume = "104",
    number = "5",
    pages = "055017",
    year = "2021"
}

@article{Alonso:2016oyd,
    author = "Alonso, Rodrigo and Grinstein, Benjam\'\i{}n and Martin Camalich, Jorge",
    title = "{Lifetime of $B_c^-$ Constrains Explanations for Anomalies in  $B\to D^{(*)}\tau\nu$}",
    eprint = "1611.06676",
    archivePrefix = "arXiv",
    primaryClass = "hep-ph",
    reportNumber = "CERN-TH-2016-230",
    doi = "10.1103/PhysRevLett.118.081802",
    journal = "Phys. Rev. Lett.",
    volume = "118",
    number = "8",
    pages = "081802",
    year = "2017"
}

@article{Akeroyd:2017mhr,
    author = "Akeroyd, A. G. and Chen, Chuan-Hung",
    title = "{Constraint on the branching ratio of $B_c \to \tau \bar{\nu}$ from LEP1 and consequences for $R(D^{(*)})$ anomaly}",
    eprint = "1708.04072",
    archivePrefix = "arXiv",
    primaryClass = "hep-ph",
    doi = "10.1103/PhysRevD.96.075011",
    journal = "Phys. Rev. D",
    volume = "96",
    number = "7",
    pages = "075011",
    year = "2017"
}

@inproceedings{Belle:2019ewo,
    author = "Abdesselam, A. and others",
    collaboration = "Belle",
    title = "{Measurement of the $D^{\ast-}$ polarization in the decay $B^0 \to D^{\ast -}\tau^+\nu_{\tau}$}",
    booktitle = "{10th International Workshop on the CKM Unitarity Triangle}",
    eprint = "1903.03102",
    archivePrefix = "arXiv",
    primaryClass = "hep-ex",
    reportNumber = "BELLE-CONF-1805",
    month = "3",
    year = "2019"
}

@online{belle2talk,
  author    = {Author, A.},
  title     = {Title of the Talk},
  year      = {2023},
  url       = {https://docs.belle2.org/record/3746/files/BELLE2-TALK-CONF-2023-097.pdf?version=2},
  note      = {Belle2 Collaboration Talk},
}

@article{BaBar:2012obs,
    author = "Lees, J. P. and others",
    collaboration = "BaBar",
    title = "{Evidence for an excess of $\bar{B} \to D^{(*)} \tau^-\bar{\nu}_\tau$ decays}",
    eprint = "1205.5442",
    archivePrefix = "arXiv",
    primaryClass = "hep-ex",
    reportNumber = "BABAR-PUB-12-012, SLAC-PUB-15028",
    doi = "10.1103/PhysRevLett.109.101802",
    journal = "Phys. Rev. Lett.",
    volume = "109",
    pages = "101802",
    year = "2012"
}

@article{BaBar:2013mob,
    author = "Lees, J. P. and others",
    collaboration = "BaBar",
    title = "{Measurement of an Excess of $\bar{B} \to D^{(*)}\tau^- \bar{\nu}_\tau$ Decays and Implications for Charged Higgs Bosons}",
    eprint = "1303.0571",
    archivePrefix = "arXiv",
    primaryClass = "hep-ex",
    reportNumber = "BABAR-PUB-13-001, SLAC-PUB-15381",
    doi = "10.1103/PhysRevD.88.072012",
    journal = "Phys. Rev. D",
    volume = "88",
    number = "7",
    pages = "072012",
    year = "2013"
}

@article{Belle:2015qfa,
    author = "Huschle, M. and others",
    collaboration = "Belle",
    title = "{Measurement of the branching ratio of $\bar{B} \to D^{(\ast)} \tau^- \bar{\nu}_\tau$ relative to $\bar{B} \to D^{(\ast)} \ell^- \bar{\nu}_\ell$ decays with hadronic tagging at Belle}",
    eprint = "1507.03233",
    archivePrefix = "arXiv",
    primaryClass = "hep-ex",
    reportNumber = "KEK-REPORT-2015-18",
    doi = "10.1103/PhysRevD.92.072014",
    journal = "Phys. Rev. D",
    volume = "92",
    number = "7",
    pages = "072014",
    year = "2015"
}

@article{Belle:2016dyj,
    author = "Hirose, S. and others",
    collaboration = "Belle",
    title = "{Measurement of the $\tau$ lepton polarization and $R(D^*)$ in the decay $\bar{B} \to D^* \tau^- \bar{\nu}_\tau$}",
    eprint = "1612.00529",
    archivePrefix = "arXiv",
    primaryClass = "hep-ex",
    reportNumber = "KEK-PREPRINT-2016-53, BELLE-PREPRINT-2016-14",
    doi = "10.1103/PhysRevLett.118.211801",
    journal = "Phys. Rev. Lett.",
    volume = "118",
    number = "21",
    pages = "211801",
    year = "2017"
}

@article{Belle:2017ilt,
    author = "Hirose, S. and others",
    collaboration = "Belle",
    title = "{Measurement of the $\tau$ lepton polarization and $R(D^*)$ in the decay $\bar{B} \rightarrow D^* \tau^- \bar{\nu}_\tau$ with one-prong hadronic $\tau$ decays at Belle}",
    eprint = "1709.00129",
    archivePrefix = "arXiv",
    primaryClass = "hep-ex",
    reportNumber = "KEK-PREPRINT-2017-26, BELLE-PREPRINT-2017-18",
    doi = "10.1103/PhysRevD.97.012004",
    journal = "Phys. Rev. D",
    volume = "97",
    number = "1",
    pages = "012004",
    year = "2018"
}

@article{Belle:2019gij,
    author = "Abdesselam, A. and others",
    collaboration = "Belle",
    title = "{Measurement of $\mathcal{R}(D)$ and $\mathcal{R}(D^{\ast})$ with a semileptonic tagging method}",
    eprint = "1904.08794",
    archivePrefix = "arXiv",
    primaryClass = "hep-ex",
    month = "4",
    year = "2019"
}

@article{Belle:2019rba,
    author = "Caria, G. and others",
    collaboration = "Belle",
    title = "{Measurement of $\mathcal{R}(D)$ and $\mathcal{R}(D^*)$ with a semileptonic tagging method}",
    eprint = "1910.05864",
    archivePrefix = "arXiv",
    primaryClass = "hep-ex",
    reportNumber = "Belle-2019-18, KEK-2019-40",
    doi = "10.1103/PhysRevLett.124.161803",
    journal = "Phys. Rev. Lett.",
    volume = "124",
    number = "16",
    pages = "161803",
    year = "2020"
}

@article{LHCb:2023zxo,
    collaboration = "LHCb",
    title = "{Measurement of the ratios of branching fractions $\mathcal{R}(D^{*})$ and $\mathcal{R}(D^{0})$}",
    eprint = "2302.02886",
    archivePrefix = "arXiv",
    primaryClass = "hep-ex",
    reportNumber = "LHCb-PAPER-2022-039, CERN-EP-2022-284",
    doi = "10.1103/PhysRevLett.131.111802",
    journal = "Phys. Rev. Lett.",
    volume = "131",
    pages = "111802",
    year = "2023"
}

@article{LHCb:2023uiv,
    author = "Aaij, Roel and others",
    collaboration = "LHCb",
    title = "{Test of lepton flavor universality using B0\textrightarrow{}D*-\ensuremath{\tau}+\ensuremath{\nu}\ensuremath{\tau} decays with hadronic \ensuremath{\tau} channels}",
    eprint = "2305.01463",
    archivePrefix = "arXiv",
    primaryClass = "hep-ex",
    reportNumber = "LHCb-PAPER-2022-052, CERN-EP-2023-062",
    doi = "10.1103/PhysRevD.108.012018",
    journal = "Phys. Rev. D",
    volume = "108",
    number = "1",
    pages = "012018",
    year = "2023"
}

@article{Iguro:2022yzr,
    author = "Iguro, Syuhei and Kitahara, Teppei and Watanabe, Ryoutaro",
    title = "{Global fit to $b \to c\tau\nu$ anomalies 2022 mid-autumn}",
    eprint = "2210.10751",
    archivePrefix = "arXiv",
    primaryClass = "hep-ph",
    reportNumber = "P3H-22-103, TTP22-062, KEK-TH-2464",
    month = "10",
    year = "2022"
}

@article{Capdevila:2023yhq,
    author = "Capdevila, Bernat and Crivellin, Andreas and Matias, Joaquim",
    title = "{Review of Semileptonic $B$ Anomalies}",
    eprint = "2309.01311",
    archivePrefix = "arXiv",
    primaryClass = "hep-ph",
    reportNumber = "PSI-PR-23-33, ZU-TH 50/23",
    doi = "10.1140/epjs/s11734-023-01012-2",
    journal = "Eur. Phys. J. ST",
    volume = "1",
    pages = "20",
    year = "2023"
}

@article{Gripaios:2009dq,
    author = "Gripaios, Ben",
    title = "{Composite Leptoquarks at the LHC}",
    eprint = "0910.1789",
    archivePrefix = "arXiv",
    primaryClass = "hep-ph",
    reportNumber = "CERN-PH-TH-2009-190",
    doi = "10.1007/JHEP02(2010)045",
    journal = "JHEP",
    volume = "02",
    pages = "045",
    year = "2010"
}

@article{Gripaios:2014tna,
    author = "Gripaios, Ben and Nardecchia, Marco and Renner, S. A.",
    title = "{Composite leptoquarks and anomalies in $B$-meson decays}",
    eprint = "1412.1791",
    archivePrefix = "arXiv",
    primaryClass = "hep-ph",
    doi = "10.1007/JHEP05(2015)006",
    journal = "JHEP",
    volume = "05",
    pages = "006",
    year = "2015"
}

@article{deMedeirosVarzielas:2015yxm,
    author = "de Medeiros Varzielas, Ivo and Hiller, Gudrun",
    title = "{Clues for flavor from rare lepton and quark decays}",
    eprint = "1503.01084",
    archivePrefix = "arXiv",
    primaryClass = "hep-ph",
    reportNumber = "DO-TH-15-02, QFET-2015-04",
    doi = "10.1007/JHEP06(2015)072",
    journal = "JHEP",
    volume = "06",
    pages = "072",
    year = "2015"
}

@article{Crivellin:2017zlb,
    author = {Crivellin, Andreas and M\"uller, Dario and Ota, Toshihiko},
    title = "{Simultaneous explanation of R(D$^{(*)}$) and b\textrightarrow{}s\ensuremath{\mu}$^{+}$ \ensuremath{\mu}$^{-}$: the last scalar leptoquarks standing}",
    eprint = "1703.09226",
    archivePrefix = "arXiv",
    primaryClass = "hep-ph",
    reportNumber = "PSI-PR-17-04, YACHAY-PUB-17-03-PN, ZU-TH-05-17",
    doi = "10.1007/JHEP09(2017)040",
    journal = "JHEP",
    volume = "09",
    pages = "040",
    year = "2017"
}

@article{Sahoo:2015wya,
    author = "Sahoo, Suchismita and Mohanta, Rukmani",
    title = "{Scalar leptoquarks and the rare $B$ meson decays}",
    eprint = "1501.05193",
    archivePrefix = "arXiv",
    primaryClass = "hep-ph",
    doi = "10.1103/PhysRevD.91.094019",
    journal = "Phys. Rev. D",
    volume = "91",
    number = "9",
    pages = "094019",
    year = "2015"
}

@article{Alonso:2015sja,
    author = "Alonso, Rodrigo and Grinstein, Benjam\'\i{}n and Martin Camalich, Jorge",
    title = "{Lepton universality violation and lepton flavor conservation in $B$-meson decays}",
    eprint = "1505.05164",
    archivePrefix = "arXiv",
    primaryClass = "hep-ph",
    doi = "10.1007/JHEP10(2015)184",
    journal = "JHEP",
    volume = "10",
    pages = "184",
    year = "2015"
}

@article{Barbieri:2015yvd,
    author = "Barbieri, Riccardo and Isidori, Gino and Pattori, Andrea and Senia, Fabrizio",
    title = "{Anomalies in $B$-decays and $U(2)$ flavour symmetry}",
    eprint = "1512.01560",
    archivePrefix = "arXiv",
    primaryClass = "hep-ph",
    reportNumber = "ZU-TH-44-15",
    doi = "10.1140/epjc/s10052-016-3905-3",
    journal = "Eur. Phys. J. C",
    volume = "76",
    number = "2",
    pages = "67",
    year = "2016"
}

@article{Becirevic:2016yqi,
    author = "Be\v{c}irevi\'c, Damir and Fajfer, Svjetlana and Ko\v{s}nik, Nejc and Sumensari, Olcyr",
    title = "{Leptoquark model to explain the $B$-physics anomalies, $R_K$ and $R_D$}",
    eprint = "1608.08501",
    archivePrefix = "arXiv",
    primaryClass = "hep-ph",
    reportNumber = "LPT-ORSAY-16-51",
    doi = "10.1103/PhysRevD.94.115021",
    journal = "Phys. Rev. D",
    volume = "94",
    number = "11",
    pages = "115021",
    year = "2016"
}

@article{Cai:2017wry,
    author = "Cai, Yi and Gargalionis, John and Schmidt, Michael A. and Volkas, Raymond R.",
    title = "{Reconsidering the One Leptoquark solution: flavor anomalies and neutrino mass}",
    eprint = "1704.05849",
    archivePrefix = "arXiv",
    primaryClass = "hep-ph",
    doi = "10.1007/JHEP10(2017)047",
    journal = "JHEP",
    volume = "10",
    pages = "047",
    year = "2017"
}

@article{Sahoo:2016pet,
    author = "Sahoo, Suchismita and Mohanta, Rukmani and Giri, Anjan K.",
    title = "{Explaining the $R_{K}$ and $R_{D^{(*)}}$ anomalies with vector leptoquarks}",
    eprint = "1609.04367",
    archivePrefix = "arXiv",
    primaryClass = "hep-ph",
    doi = "10.1103/PhysRevD.95.035027",
    journal = "Phys. Rev. D",
    volume = "95",
    number = "3",
    pages = "035027",
    year = "2017"
}

@article{Barbieri:2016las,
    author = "Barbieri, Riccardo and Murphy, Christopher W. and Senia, Fabrizio",
    title = "{B-decay Anomalies in a Composite Leptoquark Model}",
    eprint = "1611.04930",
    archivePrefix = "arXiv",
    primaryClass = "hep-ph",
    doi = "10.1140/epjc/s10052-016-4578-7",
    journal = "Eur. Phys. J. C",
    volume = "77",
    number = "1",
    pages = "8",
    year = "2017"
}

@article{Buttazzo:2017ixm,
    author = "Buttazzo, Dario and Greljo, Admir and Isidori, Gino and Marzocca, David",
    title = "{B-physics anomalies: a guide to combined explanations}",
    eprint = "1706.07808",
    archivePrefix = "arXiv",
    primaryClass = "hep-ph",
    reportNumber = "ZU-TH-18-17",
    doi = "10.1007/JHEP11(2017)044",
    journal = "JHEP",
    volume = "11",
    pages = "044",
    year = "2017"
}

@article{Marzocca:2018wcf,
    author = "Marzocca, David",
    title = "{Addressing the B-physics anomalies in a fundamental Composite Higgs Model}",
    eprint = "1803.10972",
    archivePrefix = "arXiv",
    primaryClass = "hep-ph",
    doi = "10.1007/JHEP07(2018)121",
    journal = "JHEP",
    volume = "07",
    pages = "121",
    year = "2018"
}

@article{Becirevic:2018afm,
    author = "Be\v{c}irevi\'c, Damir and Dor\v{s}ner, Ilja and Fajfer, Svjetlana and Ko\v{s}nik, Nejc and Faroughy, Darius A. and Sumensari, Olcyr",
    title = "{Scalar leptoquarks from grand unified theories to accommodate the $B$-physics anomalies}",
    eprint = "1806.05689",
    archivePrefix = "arXiv",
    primaryClass = "hep-ph",
    reportNumber = "LPT-Orsay-18-76, LPT-ORSAY-18-76",
    doi = "10.1103/PhysRevD.98.055003",
    journal = "Phys. Rev. D",
    volume = "98",
    number = "5",
    pages = "055003",
    year = "2018"
}

@article{Vignaroli:2018lpq,
    author = "Vignaroli, Natascia",
    title = "{Seeking leptoquarks in the $\bf t\bar{t}$ plus missing energy channel at the high-luminosity LHC}",
    eprint = "1808.10309",
    archivePrefix = "arXiv",
    primaryClass = "hep-ph",
    doi = "10.1103/PhysRevD.99.035021",
    journal = "Phys. Rev. D",
    volume = "99",
    number = "3",
    pages = "035021",
    year = "2019"
}

@article{Vignaroli:2019lkg,
    author = "Vignaroli, Natascia",
    editor = "D'Ambrosio, G. and De Nardo, G.",
    title = "{Leptoquarks in $B$-meson anomalies: simplified models and HL-LHC discovery prospects}",
    eprint = "1912.00899",
    archivePrefix = "arXiv",
    primaryClass = "hep-ph",
    doi = "10.1393/ncc/i2020-20053-0",
    journal = "Nuovo Cim. C",
    volume = "43",
    number = "2-3",
    pages = "53",
    year = "2020"
}

@article{BhupalDev:2020zcy,
    author = "Bhupal Dev, P. S. and Mohanta, Rukmani and Patra, Sudhanwa and Sahoo, Suchismita",
    title = "{Unified explanation of flavor anomalies, radiative neutrino masses, and ANITA anomalous events in a vector leptoquark model}",
    eprint = "2004.09464",
    archivePrefix = "arXiv",
    primaryClass = "hep-ph",
    doi = "10.1103/PhysRevD.102.095012",
    journal = "Phys. Rev. D",
    volume = "102",
    number = "9",
    pages = "095012",
    year = "2020"
}

@article{Crivellin:2022mff,
    author = "Crivellin, Andreas and Fuks, Benjamin and Schnell, Luc",
    title = "{Explaining the hints for lepton flavour universality violation with three S$_{2}$ leptoquark generations}",
    eprint = "2203.10111",
    archivePrefix = "arXiv",
    primaryClass = "hep-ph",
    reportNumber = "MPP-2022-30, PSI-PR-22-07, ZU-TH 08/22",
    doi = "10.1007/JHEP06(2022)169",
    journal = "JHEP",
    volume = "06",
    pages = "169",
    year = "2022"
}

@article{Angelescu:2018tyl,
    author = "Angelescu, A. and Be\v{c}irevi\'c, Damir and Faroughy, D. A. and Sumensari, O.",
    title = "{Closing the window on single leptoquark solutions to the $B$-physics anomalies}",
    eprint = "1808.08179",
    archivePrefix = "arXiv",
    primaryClass = "hep-ph",
    reportNumber = "LPT-Orsay-18-81",
    doi = "10.1007/JHEP10(2018)183",
    journal = "JHEP",
    volume = "10",
    pages = "183",
    year = "2018"
}

@article{Allwicher:2023syp,
    author = "Allwicher, Lukas and Becirevic, Damir and Piazza, Gioacchino and Rosauro-Alcaraz, Salvador and Sumensari, Olcyr",
    title = "{Understanding the first measurement of $\mathcal{B}(B\to K \nu \bar{\nu})$}",
    eprint = "2309.02246",
    archivePrefix = "arXiv",
    primaryClass = "hep-ph",
    month = "9",
    year = "2023"
}

@article{LHCb:2015svh,
    author = "Aaij, Roel and others",
    collaboration = "LHCb",
    title = "{Angular analysis of the $B^{0} \to K^{*0} \mu^{+} \mu^{-}$ decay using 3 fb$^{-1}$ of integrated luminosity}",
    eprint = "1512.04442",
    archivePrefix = "arXiv",
    primaryClass = "hep-ex",
    reportNumber = "CERN-PH-EP-2015-314, LHCB-PAPER-2015-051",
    doi = "10.1007/JHEP02(2016)104",
    journal = "JHEP",
    volume = "02",
    pages = "104",
    year = "2016"
}

@article{Belle:2015pkj,
    author = "Glattauer, R. and others",
    collaboration = "Belle",
    title = "{Measurement of the decay $B\to D\ell\nu_\ell$ in fully reconstructed events and determination of the Cabibbo-Kobayashi-Maskawa matrix element $|V_{cb}|$}",
    eprint = "1510.03657",
    archivePrefix = "arXiv",
    primaryClass = "hep-ex",
    reportNumber = "BELLE-PREPRINT-2015-16, KEK-PREPRINT-2015-43",
    doi = "10.1103/PhysRevD.93.032006",
    journal = "Phys. Rev. D",
    volume = "93",
    number = "3",
    pages = "032006",
    year = "2016"
}

@article{BaBar:2009zxk,
    author = "Aubert, Bernard and others",
    collaboration = "BaBar",
    title = "{Measurement of |V(cb)| and the Form-Factor Slope in anti-B ---\ensuremath{>} D l- anti-nu Decays in Events Tagged by a Fully Reconstructed B Meson}",
    eprint = "0904.4063",
    archivePrefix = "arXiv",
    primaryClass = "hep-ex",
    reportNumber = "BABAR-PUB-09-009, SLAC-PUB-13580",
    doi = "10.1103/PhysRevLett.104.011802",
    journal = "Phys. Rev. Lett.",
    volume = "104",
    pages = "011802",
    year = "2010"
}

@article{Belle:2010qug,
    author = "Dungel, W. and others",
    collaboration = "Belle",
    title = "{Measurement of the form factors of the decay B0 -\ensuremath{>} D*- ell+ nu and determination of the CKM matrix element |Vcb|}",
    eprint = "1010.5620",
    archivePrefix = "arXiv",
    primaryClass = "hep-ex",
    doi = "10.1103/PhysRevD.82.112007",
    journal = "Phys. Rev. D",
    volume = "82",
    pages = "112007",
    year = "2010"
}

@article{Belle:2017rcc,
    author = "Abdesselam, A. and others",
    collaboration = "Belle",
    title = "{Precise determination of the CKM matrix element $\left| V_{cb}\right|$ with $\bar B^0 \to D^{*\,+} \, \ell^- \, \bar \nu_\ell$ decays with hadronic tagging at Belle}",
    eprint = "1702.01521",
    archivePrefix = "arXiv",
    primaryClass = "hep-ex",
    reportNumber = "BELLE-CONF-1612",
    month = "2",
    year = "2017"
}

@inproceedings{Adamczyk:2019wyt,
    author = "Adamczyk, Karol",
    collaboration = "Belle, Belle-II",
    title = "{Semitauonic $B$ decays at Belle/Belle II}",
    booktitle = "{10th International Workshop on the CKM Unitarity Triangle}",
    eprint = "1901.06380",
    archivePrefix = "arXiv",
    primaryClass = "hep-ex",
    month = "1",
    year = "2019"
}

@article{HFLAV:2016hnz,
    author = "Amhis, Y. and others",
    collaboration = "HFLAV",
    title = "{Averages of $b$-hadron, $c$-hadron, and $\tau$-lepton properties as of summer 2016}",
    eprint = "1612.07233",
    archivePrefix = "arXiv",
    primaryClass = "hep-ex",
    reportNumber = "FERMILAB-PUB-16-611-ND",
    doi = "10.1140/epjc/s10052-017-5058-4",
    journal = "Eur. Phys. J. C",
    volume = "77",
    number = "12",
    pages = "895",
    year = "2017"
}

@article{BaBar:2008zui,
    author = "Aubert, Bernard and others",
    collaboration = "BaBar",
    title = "{Measurements of the Semileptonic Decays anti-B ---\ensuremath{>} D l anti-nu and anti-B ---\ensuremath{>} D* l anti-nu Using a Global Fit to D X l anti-nu Final States}",
    eprint = "0809.0828",
    archivePrefix = "arXiv",
    primaryClass = "hep-ex",
    reportNumber = "SLAC-PUB-13371, BABAR-PUB-08-027",
    doi = "10.1103/PhysRevD.79.012002",
    journal = "Phys. Rev. D",
    volume = "79",
    pages = "012002",
    year = "2009"
}

@article{BaBar:2007cke,
    author = "Aubert, Bernard and others",
    collaboration = "BaBar",
    title = "{Determination of the form-factors for the decay $B^0 \to D^{*-} \ell^{+} \nu_{l}$ and of the CKM matrix element $|V_{cb}|$}",
    eprint = "0705.4008",
    archivePrefix = "arXiv",
    primaryClass = "hep-ex",
    reportNumber = "SLAC-PUB-12511, BABAR-PUB-07-008",
    doi = "10.1103/PhysRevD.77.032002",
    journal = "Phys. Rev. D",
    volume = "77",
    pages = "032002",
    year = "2008"
}

@article{BaBar:2007ddh,
    author = "Aubert, Bernard and others",
    collaboration = "BaBar",
    title = "{A Measurement of the branching fractions of exclusive $\bar{B} \to D^{(*)}$ ($\pi$) $\ell^{-} \bar{\nu}$( $\ell^{)}$ decays in events with a fully reconstructed $B$ meson}",
    eprint = "0712.3503",
    archivePrefix = "arXiv",
    primaryClass = "hep-ex",
    reportNumber = "SLAC-PUB-13056, BABAR-PUB-07-071",
    doi = "10.1103/PhysRevLett.100.151802",
    journal = "Phys. Rev. Lett.",
    volume = "100",
    pages = "151802",
    year = "2008"
}

@article{BaBar:2007nwi,
    author = "Aubert, Bernard and others",
    collaboration = "BaBar",
    title = "{Measurement of the Decay $B^{-} \to$ D*0 $e^{-} \bar{\nu}$( $e$)}",
    eprint = "0712.3493",
    archivePrefix = "arXiv",
    primaryClass = "hep-ex",
    reportNumber = "SLAC-PUB-13035, BABAR-PUB-07-070",
    doi = "10.1103/PhysRevLett.100.231803",
    journal = "Phys. Rev. Lett.",
    volume = "100",
    pages = "231803",
    year = "2008"
}

@article{1310.1082,
    author = "Gauld, Rhorry and Goertz, Florian and Haisch, Ulrich",
    title = "{An explicit Z'-boson explanation of the $B \to K^* \mu^+ \mu^-$ anomaly}",
    eprint = "1310.1082",
    archivePrefix = "arXiv",
    primaryClass = "hep-ph",
    doi = "10.1007/JHEP01(2014)069",
    journal = "JHEP",
    volume = "01",
    pages = "069",
    year = "2014"
}

@article{1311.6729,
    author = "Buras, Andrzej J. and De Fazio, Fulvia and Girrbach, Jennifer",
    title = "{331 models facing new $b \to s\mu^+ \mu^-$ data}",
    eprint = "1311.6729",
    archivePrefix = "arXiv",
    primaryClass = "hep-ph",
    reportNumber = "FLAVOUR(267104)-ERC-55, BARI-TH-13-681",
    doi = "10.1007/JHEP02(2014)112",
    journal = "JHEP",
    volume = "02",
    pages = "112",
    year = "2014"
}

@article{1403.1269,
    author = "Altmannshofer, Wolfgang and Gori, Stefania and Pospelov, Maxim and Yavin, Itay",
    title = "{Quark flavor transitions in $L_\mu-L_\tau$ models}",
    eprint = "1403.1269",
    archivePrefix = "arXiv",
    primaryClass = "hep-ph",
    doi = "10.1103/PhysRevD.89.095033",
    journal = "Phys. Rev. D",
    volume = "89",
    pages = "095033",
    year = "2014"
}

@article{1501.00993,
    author = "Crivellin, Andreas and D'Ambrosio, Giancarlo and Heeck, Julian",
    title = "{Explaining $h\to\mu^\pm\tau^\mp$, $B\to K^* \mu^+\mu^-$ and $B\to K \mu^+\mu^-/B\to K e^+e^-$ in a two-Higgs-doublet model with gauged $L_\mu-L_\tau$}",
    eprint = "1501.00993",
    archivePrefix = "arXiv",
    primaryClass = "hep-ph",
    reportNumber = "CERN-PH-TH-2015-001, ULB-TH-14-26",
    doi = "10.1103/PhysRevLett.114.151801",
    journal = "Phys. Rev. Lett.",
    volume = "114",
    pages = "151801",
    year = "2015"
}

@article{1503.03477,
    author = "Crivellin, Andreas and D'Ambrosio, Giancarlo and Heeck, Julian",
    title = "{Addressing the LHC flavor anomalies with horizontal gauge symmetries}",
    eprint = "1503.03477",
    archivePrefix = "arXiv",
    primaryClass = "hep-ph",
    reportNumber = "CERN-PH-TH-2015-046, ULB-TH-15-03",
    doi = "10.1103/PhysRevD.91.075006",
    journal = "Phys. Rev. D",
    volume = "91",
    number = "7",
    pages = "075006",
    year = "2015"
}

@article{1503.03865,
    author = "Niehoff, Christoph and Stangl, Peter and Straub, David M.",
    title = "{Violation of lepton flavour universality in composite Higgs models}",
    eprint = "1503.03865",
    archivePrefix = "arXiv",
    primaryClass = "hep-ph",
    doi = "10.1016/j.physletb.2015.05.063",
    journal = "Phys. Lett. B",
    volume = "747",
    pages = "182--186",
    year = "2015"
}

@article{1505.03079,
    author = "Celis, Alejandro and Fuentes-Martin, Javier and Jung, Martin and Serodio, Hugo",
    title = "{Family nonuniversal $Z'$ models with protected flavor-changing interactions}",
    eprint = "1505.03079",
    archivePrefix = "arXiv",
    primaryClass = "hep-ph",
    reportNumber = "LMU-ASC-22-15, IFIC-15-23, FLAVOUR(267104)-ERC-95",
    doi = "10.1103/PhysRevD.92.015007",
    journal = "Phys. Rev. D",
    volume = "92",
    number = "1",
    pages = "015007",
    year = "2015"
}

@article{1506.01705,
    author = "Greljo, Admir and Isidori, Gino and Marzocca, David",
    title = "{On the breaking of Lepton Flavor Universality in B decays}",
    eprint = "1506.01705",
    archivePrefix = "arXiv",
    primaryClass = "hep-ph",
    reportNumber = "ZU-TH-16-15",
    doi = "10.1007/JHEP07(2015)142",
    journal = "JHEP",
    volume = "07",
    pages = "142",
    year = "2015"
}

@article{1508.07009,
    author = "Altmannshofer, Wolfgang and Yavin, Itay",
    title = "{Predictions for lepton flavor universality violation in rare B decays in models with gauged $L_\mu - L_\tau$}",
    eprint = "1508.07009",
    archivePrefix = "arXiv",
    primaryClass = "hep-ph",
    doi = "10.1103/PhysRevD.92.075022",
    journal = "Phys. Rev. D",
    volume = "92",
    number = "7",
    pages = "075022",
    year = "2015"
}

@article{1509.01249,
    author = "Falkowski, Adam and Nardecchia, Marco and Ziegler, Robert",
    title = "{Lepton Flavor Non-Universality in B-meson Decays from a U(2) Flavor Model}",
    eprint = "1509.01249",
    archivePrefix = "arXiv",
    primaryClass = "hep-ph",
    doi = "10.1007/JHEP11(2015)173",
    journal = "JHEP",
    volume = "11",
    pages = "173",
    year = "2015"
}

@article{1510.07658,
    author = "Carmona, Adrian and Goertz, Florian",
    title = "{Lepton Flavor and Nonuniversality from Minimal Composite Higgs Setups}",
    eprint = "1510.07658",
    archivePrefix = "arXiv",
    primaryClass = "hep-ph",
    reportNumber = "CERN-PH-TH-2015-251",
    doi = "10.1103/PhysRevLett.116.251801",
    journal = "Phys. Rev. Lett.",
    volume = "116",
    number = "25",
    pages = "251801",
    year = "2016"
}

@article{1512.08500,
    author = "Goertz, Florian and Kamenik, Jernej F. and Katz, Andrey and Nardecchia, Marco",
    title = "{Indirect Constraints on the Scalar Di-Photon Resonance at the LHC}",
    eprint = "1512.08500",
    archivePrefix = "arXiv",
    primaryClass = "hep-ph",
    reportNumber = "CERN-PH-TH-2015-313",
    doi = "10.1007/JHEP05(2016)187",
    journal = "JHEP",
    volume = "05",
    pages = "187",
    year = "2016"
}

@article{1601.07328,
    author = "Chiang, Cheng-Wei and He, Xiao-Gang and Valencia, German",
    title = "{$Z'$ model for $b\rightarrow s$ \ensuremath{\ell}$\overline{\ell}$ flavor anomalies}",
    eprint = "1601.07328",
    archivePrefix = "arXiv",
    primaryClass = "hep-ph",
    reportNumber = "COEPP-MN-16-1",
    doi = "10.1103/PhysRevD.93.074003",
    journal = "Phys. Rev. D",
    volume = "93",
    number = "7",
    pages = "074003",
    year = "2016"
}

@article{1602.00881,
    author = "Be\v{c}irevi\'c, Damir and Sumensari, Olcyr and Zukanovich Funchal, Renata",
    title = "{Lepton flavor violation in exclusive $b\rightarrow s$ decays}",
    eprint = "1602.00881",
    archivePrefix = "arXiv",
    primaryClass = "hep-ph",
    reportNumber = "LPT-15-57",
    doi = "10.1140/epjc/s10052-016-3985-0",
    journal = "Eur. Phys. J. C",
    volume = "76",
    number = "3",
    pages = "134",
    year = "2016"
}

@article{1604.03088,
    author = "Boucenna, Sofiane M. and Celis, Alejandro and Fuentes-Martin, Javier and Vicente, Avelino and Virto, Javier",
    title = "{Non-abelian gauge extensions for B-decay anomalies}",
    eprint = "1604.03088",
    archivePrefix = "arXiv",
    primaryClass = "hep-ph",
    reportNumber = "SI-HEP-2016-10, QFET-2016-05, LMU-ASC-16-16, IFIC-16-19",
    doi = "10.1016/j.physletb.2016.06.067",
    journal = "Phys. Lett. B",
    volume = "760",
    pages = "214--219",
    year = "2016"
}

@article{1608.01349,
    author = "Boucenna, Sofiane M. and Celis, Alejandro and Fuentes-Martin, Javier and Vicente, Avelino and Virto, Javier",
    title = "{Phenomenology of an $SU(2) \times SU(2) \times U(1)$ model with lepton-flavour non-universality}",
    eprint = "1608.01349",
    archivePrefix = "arXiv",
    primaryClass = "hep-ph",
    reportNumber = "LMU-ASC-34-16, IFIC-16-62",
    doi = "10.1007/JHEP12(2016)059",
    journal = "JHEP",
    volume = "12",
    pages = "059",
    year = "2016"
}

@article{1608.02362,
    author = "Megias, Eugenio and Panico, Giuliano and Pujolas, Oriol and Quiros, Mariano",
    title = "{A Natural origin for the LHCb anomalies}",
    eprint = "1608.02362",
    archivePrefix = "arXiv",
    primaryClass = "hep-ph",
    reportNumber = "MPP-2016-159, UAB-FT-770",
    doi = "10.1007/JHEP09(2016)118",
    journal = "JHEP",
    volume = "09",
    pages = "118",
    year = "2016"
}

@article{1611.03507,
    author = "Garcia Garcia, Isabel",
    title = "{LHCb anomalies from a natural perspective}",
    eprint = "1611.03507",
    archivePrefix = "arXiv",
    primaryClass = "hep-ph",
    doi = "10.1007/JHEP03(2017)040",
    journal = "JHEP",
    volume = "03",
    pages = "040",
    year = "2017"
}

@article{1702.08666,
    author = "Ko, P. and Omura, Yuji and Shigekami, Yoshihiro and Yu, Chaehyun",
    title = "{LHCb anomaly and B physics in flavored Z' models with flavored Higgs doublets}",
    eprint = "1702.08666",
    archivePrefix = "arXiv",
    primaryClass = "hep-ph",
    doi = "10.1103/PhysRevD.95.115040",
    journal = "Phys. Rev. D",
    volume = "95",
    number = "11",
    pages = "115040",
    year = "2017"
}

@article{1703.06019,
    author = "Megias, Eugenio and Quiros, Mariano and Salas, Lindber",
    title = "{Lepton-flavor universality violation in R$_{K}$ and $ {R}_{D^{{\left(\ast \right)}}} $ from warped space}",
    eprint = "1703.06019",
    archivePrefix = "arXiv",
    primaryClass = "hep-ph",
    reportNumber = "UAB-FT-773",
    doi = "10.1007/JHEP07(2017)102",
    journal = "JHEP",
    volume = "07",
    pages = "102",
    year = "2017"
}
\end{document}